\documentclass[10pt]{article}
\usepackage{amssymb}
\usepackage{amsmath}
\include{diagxy}
\usepackage[utf8]{inputenc}
\usepackage[T1]{fontenc}
 \linethickness{0.4pt}

\begin{document}
 \newtheorem {proposition}{Proposition}[section]
 \newtheorem{lemma}{Lemma}[section]
 \newtheorem{theorem}{Theorem}[section]
 \newtheorem{corollary}{Corollary}[section]
 \newtheorem{definition}{Definition}[section]
 \newtheorem{remark}{Remark}[section]

 \numberwithin{equation}{section}

\bigskip
\bigskip
\begin{center}
{\large\bf SOME ASPECTS OF POSITIVE KERNEL METHOD\\ OF QUANTIZATION}
\end{center}
\bigskip
\bigskip
\begin{center}
{\bf Anatol Odzijewicz\footnote{ E-mail: aodzijew@uwb.edu.pl},
Maciej Horowski}\footnote{ E-mail: horowski@math.uwb.edu.pl}

\end{center}
\bigskip
\bigskip
\begin{center}
{Faculty of Mathematics \\ University of Bia{\l}ystok
\\ K. Cio{\l}kowskiego 1M, 15-245 Bia{\l}ystok, Poland}
\end{center}
\bigskip\bigskip

\begin{abstract}
\noindent We discuss various aspects of the positive kernel method of quantization of the one-parameter groups $\tau_t \in \mbox{Aut}(P,\vartheta)$ of automorphisms of a $G$-principal bundle $P(G,\pi,M)$ with a fixed connection form $\vartheta$ on its total space $P$. We show that the generator $\hat{F}$ of the unitary flow $U_t = e^{it \hat{F}}$ being the quantization of $\tau_t $ is realized by a generalized Kirillov-Kostant-Souriau operator whose domain consists of sections of some vector bundle over $M$, which are defined by a suitable positive kernel.

This method of quantization applied to the case when  $G=GL(N,\mathbb{C})$ and $M$ is a non-compact Riemann surface leads to quantization of the arbitrary holomorphic flow $\tau_t^{hol} \in \mbox{Aut}(P,\vartheta)$. For the above case,  we present the integral decompositions of the positive kernels on $P\times P$ invariant with respect to the flows $\tau_t^{hol}$ in terms of the spectral measure of $\hat{F}$. These decompositions generalize the ones given by Bochner's Theorem for the positive kernels on $\mathbb{C} \times \mathbb{C}$ invariant with respect to the one-parameter groups of translations of complex plane.
\end{abstract}

\tableofcontents

\section{Introduction}

From the very beginnings of quantum mechanics the problem of quantization is one of the most fascinating and crucial ones for understanding the correspondence between classical and quantum physics. Excluding the field theory and restricting to the  case of mechanics only, by quantization of a Hamiltonian flow $\mathbb{R}\ni t\mapsto \sigma_t^F\in \mbox{SpDiff}(M,\omega)$ defined on a symplectic manifold $(M,\omega)$ one usually understands the construction of a corresponding unitary flow $\mathbb{R}\ni t\mapsto e^{it\widehat{F}}$ on a Hilbert space ${\cal H}$. Additionally one claims that the map $Q: {\cal P}^\infty(M,\mathbb{R})\ni F\mapsto i\widehat{F}\in {\cal L}({\cal D})$ which assigns to a classical generator $F$ the quantum one $\widehat{F}$ (a self-adjoint operator in ${\cal H}$) is a morphism of some Lie algebras, where ${\cal P}^\infty(M,\mathbb{R})$ is a Lie subalgebra of the Poisson algebra $C^\infty(M,\mathbb{R})$ and ${\cal L}({\cal D})$ is a Lie algebra of  anti-self-adjoint operators having a common domain ${\cal D}$ dense in ${\cal H}$.

Among known methods of quantization the Kirillov-Kostant-Souriau geometric quantization \cite{Ki}, \cite{Ko}, \cite{S} is  one of the most elegant from a geometric point of view and gives a precise construction of the quantum generator $\widehat{F}$ for the given classical one $F\in {\cal P}^\infty(M,\mathbb{R})$. For this construction one needs to obtain a $\sigma_t^F$-invariant complex Lagrangian distribution ${\cal P}\subset T^{\mathbb{C}}M$ and the appropriate measure (density) on the quotient manifold $M/{\cal P}\cap \bar {\cal P}$. However, this leads to serious difficulties if one wants to quantize concrete mechanical systems. In order to omit these difficulties and for deeper understanding of the relationship between the classical $(M,\omega)$ and quantum $(\mathbb{C}\mathbb{P}({\cal H}),\omega_{FS})$ phase spaces in \cite{O1} and \cite {O-H} a method of quantization based on the notion of positive kernel (coherent state map) was proposed, which in our opinion completes the Kirillov-Kostant-Souriau quantization in a natural way. For example one can find the application of the coherent state method of quantization to concrete physical systems in \cite{H-O}, \cite{O-S}.

For a general theory of positive (reproducing) kernels  and its role in differential geometry (including Banach differential manifolds and vector bundles over them) and representation theory we address to \cite{B-G1,B-G2} and to the monograph \cite{N}. See also the classical paper \cite{A} of N. Aronszajn.

Basing partly on \cite{O-H}, in Section 2 and Section 3 we briefly discuss how to extend the Kirillov-Kostant-Souriau prequantization procedure defined for $U(1)$-principal bundle to the case of an arbitrary $G$-principal bundle $P(G,\pi,M)$ with a fixed connection form $\vartheta$ on the total space $P$. In Section 2 we define the Poisson $C^\infty (M, \mathbb{R})$-module ${\cal P}^\infty_G (P,\vartheta)$ of generators $(X, F) \in{\cal P}^\infty_G(P, \vartheta) $ of generalized Hamiltonian flows $\tau_t^{(X,F)} \in \mbox{Aut}(P, \vartheta)$, i.e. those which are solutions of generalized Hamilton equations (\ref{lie54}). In Section 3 we generalize the Kirillov-Kostant-Souriau prequantization morphisms to the morphism  $Q:{\cal P}^\infty_G(P,\vartheta)\rightarrow {\cal D}^1\Gamma^\infty(M,\mathbb{V})$ of $P^\infty_G(P,v)$ in the $C^\infty (M, \mathbb{R})$-module of differential operators of  order less or equal one acting on the smooth sections $\Gamma^\infty (M, \mathbb{V})$ of an associated smooth vector bundle $\mathbb{V} \rightarrow M$ over $M$.

In Section 4 we consider the $G$-equivariant coherent state map $\mathfrak{K}:P\rightarrow \mathcal{B}(V,{\cal H})$ and the positive definite $G$-equivariant kernel $K:P\times P\rightarrow {\cal B}(V)$, where $V$ and ${\cal H}$ are complex Hilbert spaces and $\mathcal{B}(V,{\cal H})$ is the right Hilbert $\mathcal{B}(V)$-module of bounded linear maps of $V$ in ${\cal H}$. In the same section the equivalence of the coherent state $\mathfrak{K}$ and positive kernel $K$ notions is shown and the method of quantization based on them is investigated. Among others we show that the Kirillov-Kostant-Souriau differential operator $Q_{(X,F)}$ can be treated as a self-adjoint operator $\hat{F}$ in the Hilbert space ${\cal H}_K$ whose domain is defined by the $G$-equivariant positive kernel $K:P\times P\rightarrow {\cal B}(V)$ (see (\ref{nnnny1}) and (\ref{n716y1})). The conditions on this kernel needed to quantize $\tau^{(X,F)}_t\in\mbox{Aut}(P,\vartheta)$ are presented in (\ref{44432f}) and (\ref{der434}).

In Section 5 assuming that $G\subset GL(V,\mathbb{C})$ is a Lie subgroup of $GL(V, \mathbb{C})$ and that there exists a coherent state map $\mathfrak{K}: P \rightarrow B(V, \mathcal{H})$ on the total space of $P(G,\pi,M)$, we define in a canonical way two other principal bundles $\widetilde{P}(GL(V,\mathbb{C}),\widetilde{\pi},M)$ and $U(U(V),\pi^u,M)$ over $M$. The connection forms $\widetilde{\vartheta}\in \Gamma^\infty(\widetilde{P}, T^*\widetilde{P}\otimes {\cal B}(V))$ and $\vartheta^a \in \Gamma^\infty (U,T^*U\otimes T_e U(V))$ as well as the respective coherent states maps $\widetilde{\mathfrak{K}}:\widetilde{P}\rightarrow B(V, \mathcal{H})$ and $\mathfrak{a}: U\rightarrow B(V, \mathcal{H})$ are defined on these principal bundles by using of $\mathfrak{K}: P \rightarrow B(V, \mathcal{H})$. Next we show, see Proposition \ref{propositiok}, that the flows $\tau^{(X, \tilde{F})}_t \in \mbox{Aut}(\widetilde{P}, \widetilde{\vartheta})$ and $\tau_t^{(X, F^a)}\in \mbox{Aut} (U, \vartheta^a)$ have the same quantum counterpart $e^{it\hat{F}}$ as the flow $\tau_t^{(X, F)} \in \mbox{Aut}(P, \vartheta )$.

In Section 6 we quantize the holomorphic one-parameter groups of automorphisms of a holomorphic principal bundles $P(GL(V,\mathbb{C}),\pi,M)$ over a non-compact Riemann surface $M$. For this relatively simple but non-trivial case the investigated theory is presented in a complete way. In particular we obtain a Bochner-type integral decompositions of the $\tau_t^{(X,F)}$-invariant positive kernels on $P\times P$   and show their relationship with the spectral decomposition of the corresponding quantum generators $Q_{(X,F)} = \hat{F}$.

Some applications of the coherent state method in physics including quantum optics are shortly discussed in Section \ref{section7}.

\section{The Poisson $C^\infty (M, \mathbb{R})$-module corresponding to the group $\mbox{Aut}(P,\vartheta)$}\label{subsec21}

The main task of this section is the investigation of some variant of Hamiltonian mechanics on a $G$-principal bundle  $P(G,\pi,M)$, where the role of the symplectic form is played by the curvature form $\Omega$ of a fixed connection form $\vartheta\in \Gamma^\infty(P, T^*P\otimes T_eG)$. We will define the Poisson $C^\infty (M, \mathbb{R})$-module $({\cal P}^\infty_G (P, \vartheta ), \{\cdot, \cdot \}_{\vartheta})$ with the Lie bracket $ \{\cdot, \cdot \}_{\vartheta}$ given in (\ref{ytyioi423spr}), which satisfies the Leibniz property in the sense of \eqref{222}. The corresponding generalization (\ref{lie54}) of the Hamiltonian equation is presented.  The model proposed here in the case when $G= U(1)$ and $\vartheta$ has non-singular curvature form reduces to standard Hamiltonian mechanics on a symplectic manifold.

Let $\mbox{Aut}(P,\vartheta)$  denote the group of diffeomorphisms of $P$ which preserve the principal bundle structure of $P$ and the connection form $\vartheta$. We recall that a $T_eG$- valued differential one-form $\vartheta$ on $P$ is a connection form  if and only if
\begin{equation}\label{kon4}
    \vartheta(pg)\circ T\kappa_g(p)=\mbox{Ad}_{g^{-1}}\circ\vartheta(p),
\end{equation}
\begin{equation}\label{kon4a}
    \vartheta(p)\circ T\kappa_p(e)=\mbox{id}_{T_eG},
\end{equation}
for any $p\in P$ and $g\in G$, where $\kappa: P\times G\rightarrow P$ is the right-action $\kappa(p,g)=:pg$ of the Lie group $G$ on $P$ and $T_eG$ is the tangent space to $G$ at the unit element $e\in G$, i.e. the Lie algebra of $G$. By $\kappa_g: P\rightarrow P$ and $\kappa_p: G\rightarrow P$ we denoted the maps defined by $\kappa_g(p):=pg$ and $\kappa_p(g):=pg$. The connection form $\vartheta$ defines the spliting
\begin{equation}\label{spl0}
    T_pP=T_p^vP\oplus T_p^hP,\nonumber
\end{equation}
of the tangent space $T_pP$  at $p\in P$ on the horizontal $T_p^hP:=\mbox{ker}\;\vartheta(p)$ and vertical $T_p^vP$, i.e. tangent to the fibre $\pi^{-1}(\pi(p))$ of $\pi:P\rightarrow M$, components (see e.g. \cite{K-N}).

Let $\tau:(\mathbb R,+)\rightarrow(\mbox{Aut}(P,\vartheta),\circ)$  be a one-parameter subgroup of $\mbox{Aut}(P,\vartheta)$ i.e. the map $\tau: \mathbb R \times P\to P$ is a smooth map which satisfies
\begin{equation}\label{juh6qq}
    \tau_t(pg)=\tau_t(p)g\;\;\;and\;\;\;\tau_t^*\vartheta=\vartheta.\nonumber
\end{equation}
Then for the vector field $\xi\in \Gamma^\infty(TP)$ tangent to the flow $\{\tau_t\}_{t\in\mathbb R}$ one has
\begin{equation}\label{pr4es}
    T\kappa_g(p)\xi(p)=\xi(pg),
 \end{equation}
\begin{equation}\label{presssd}
    {\cal L}_\xi\vartheta=0,
\end{equation}
 where ${\cal L}_\xi$ is the Lie derivative with respect to $\xi$.

 The $C^\infty(M,\mathbb{R})$-modules of vector fields $\xi\in\Gamma^\infty(TP)$ which satisfy the condition (\ref{pr4es}) and the condition (\ref{pr4es}) together with the condition (\ref{presssd}) will be denoted by $\Gamma_G^\infty(TP)$ and
 $\Gamma_{G,\vartheta}^\infty(TP)$, respectively. The $C^\infty(M,\mathbb{R})$-module structure on $\Gamma_G^\infty(TP)$, and hence on its submodules  $\Gamma_G^\infty(T^hP)$ and $\Gamma_G^\infty(T^vP)$, is defined by
 \begin{equation}\label{iomod4}
    C^\infty(M,\mathbb{R})\times\Gamma_G^\infty(TP)\ni(f,\xi) \longmapsto (f\circ\pi)\xi\in \Gamma_G^\infty(TP),\nonumber
 \end{equation}
 where by $T^hP$ and $T^vP$ we denote the horizontal and vertical vector subbundles of $TP$, respectively.

The tangent map $T\pi:TP\rightarrow TM$ defines an isomorphism of $C^\infty(M,\mathbb{R})$-modules of horizontal vector fields $\Gamma_G^\infty(T^hP)$ on $P$ and vector fields $\Gamma^\infty(TM)$ on $M$ by
 \begin{equation}\label{gwiazdki7}
    (\pi^*\xi^h)(\pi(p)):=T\pi(p)\xi^h(p).
 \end{equation}

In the sequel we will need the $C^\infty(M,\mathbb{R})$-module $C^\infty_G(P,T_eG)$ which by definition consists of smooth functions $F:P\rightarrow T_eG$ satisfying the $G$-equivariance property
 \begin{equation}\label{ioikon4e}
    F(pg)=\mbox{Ad}_{g^{-1}}F(p).\nonumber
\end{equation}
This module is isomorphic to the module of vertical vector fields $\Gamma_G^\infty(T^vP)$ on $P$, where the $C^\infty(M,\mathbb{R})$-module isomorphism $\nu^*:\Gamma_G^\infty(T^vP)\stackrel{\sim}{\longrightarrow} C^\infty_G(P,T_eG)$ is defined as follows
\begin{equation}\label{sioipr4}
    \nu^*(\xi^v)(p):=\vartheta(p)(\xi^v(p))=\vartheta(p)(\xi(p)).
\end{equation}
Its inverse is given by
\begin{equation}\label{ipoipr4}
    (\nu^*)^{-1}(F)(p)=T\kappa_p(e)F(p).\nonumber
\end{equation}
The correctness of the definitions above follows from (\ref{kon4}) and (\ref{kon4a}), and from $T\kappa_g(p)\circ T\kappa_p(e)=T\kappa_{pg}(e)$.

Taking the decomposition $\xi=\xi^h+\xi^v$ of $\xi\in \Gamma_G^\infty(TP)$ on the horizontal and vertical components and isomorphisms (\ref{gwiazdki7}) and (\ref{sioipr4}) we define
\begin{equation}\label{ioioio8}
    (X(\pi(p)), F(p))=(\pi^*\times \nu^*)(\xi)(p):=(T\pi(p)\xi^h(p),\vartheta(p)(\xi(p)))\nonumber
\end{equation}
 a $C^\infty(M,\mathbb{R})$-module isomorphism $\pi^*\times \nu^*: \Gamma_G^\infty(TP)\stackrel{\sim}{\longrightarrow}\Gamma^\infty(TM)\times C^\infty_G(P,T_eG)$.
The inverse of $\pi^*\times \nu^*$ is given by
\begin{equation}\label{ioiokj8}
    (\pi^*\times \nu^*)^{-1}(X, F)(p)=H^*(X)(p)+ T\kappa_p(e)F(p),\nonumber
\end{equation}
where by $H^*:\Gamma^\infty(TM)\rightarrow\Gamma_G^\infty(T^hP)$ we denote the horizontal lift, i.e. the module isomorphism inverse to $\pi^*$.

For $\xi\in  \Gamma_G^\infty(TP)$ one has
\begin{equation}
    {\cal L}_\xi\vartheta= {\cal L}_{\xi^h}\vartheta +{\cal L}_{\xi^v}\vartheta=d(\xi^h\llcorner\vartheta )+ \xi^h\llcorner d\vartheta+d(\xi^v\llcorner\vartheta )+ \xi^v\llcorner d\vartheta=\nonumber
\end{equation}
\begin{equation}\label{ioilie54}
    = \xi^h\llcorner \Omega+d(\xi^v\llcorner\vartheta )+ [\xi^v\llcorner \vartheta,\vartheta]= H^*(X)\llcorner \Omega+dF + [F,\vartheta]=H^*(X)\llcorner \Omega+\textbf{D}F,
\end{equation}
where
\begin{equation}\label{iostr4e}
    \Omega:=\textbf{D}\vartheta=d\vartheta+\frac{1}{2} [\vartheta,\vartheta]\nonumber
\end{equation}
is the curvature form of $\vartheta$ and
\begin{equation}\label{iokodyfu}
    \textbf{D}F=dF+[\vartheta,F]\nonumber
\end{equation}
is the covariant derivative of $F$.

On $\Gamma_G^\infty(TP)$ one has the structure of Lie algebra given by the Lie bracket $[\cdot, \cdot ]$ of vector fields.  Using the $C^\infty (M, \mathbb{R})$-modules isomorphism $\pi^*\times \nu^*: \Gamma_G^\infty(TP)\stackrel{\sim}{\longrightarrow}\Gamma^\infty(TM)\times C^\infty_G(P,T_eG),$
we can carry the Lie bracket $[\cdot, \cdot ]$ from $\Gamma_G^\infty(TP)$ to $\Gamma^\infty(TM)\times C^\infty_G(P,T_eG)$ obtaining  in this way the Lie bracket
\begin{equation}
\{(X_1,F_1),(X_2,F_2)\}_\vartheta=\nonumber
\end{equation}
\begin{equation}\label{ioi423spr}
=([X_1,X_2],-H^*(X_2))\llcorner \textbf{D}F_1 +H^*(X_1))\llcorner \textbf{D}F_2 -2\Omega(H^*(X_1),H^*(X_2))-[F_1,F_2]),
\end{equation}
of $(X_i,F_i)=(\pi^*\times \nu^*)(\xi_i)$, $i=1,2$, where $[F_1,F_2](p):=[F_1(p),F_2(p)]$.

It is reasonable to mention here that the Lie algebra $(\Gamma_G^\infty(TP), [\cdot, \cdot ])$ is isomorphic with the Lie algebra $(\Gamma^\infty (TP/G), [\cdot, \cdot ])$ of the Atiyah Lie algebroid, being a central ingredient of the Atiyah exact sequence of algebroids
$$0 \rightarrow P\times_{Ad^{-1}} T_eG \stackrel{l}{\hookrightarrow} TP/G \stackrel{a}{\rightarrow}TM \rightarrow 0,$$
see e.g. \cite{M}.

Though the language of Lie algebroid theory will not be used later, we note that the projection on the first component $\mbox{pr}_1: \Gamma^\infty (TM)\times C^\infty_G(P, T_eG) \rightarrow \Gamma^\infty (TM)$ corresponds to the anchor map $a: TP/G \rightarrow TM $ of the Atiyah algebroid. Hence, from the defining property of the anchor map we have
\begin{equation}
\mbox{pr}_1 \{(X_1,F_1),(X_2,F_2)\}_\vartheta = [\mbox{pr}_1 (X_1, F_1), \mbox{pr}_1 (X_2, F_2)]\nonumber
\end{equation}
and
\begin{equation}\label{eq:219}
\{(X_1,F_1),f(X_2,F_2)\}_\vartheta = f \{(X_1,F_1),(X_2,F_2)\}_\vartheta + X_1(f)(X_2, F_2).
\end{equation}
These properties of $\mbox{pr}_1$ and $\{\cdot, \cdot \}_{\vartheta}$ can also be obtained directly from their definitions.

The above structure of Lie $C^\infty (M)$-module on $\Gamma^\infty (TM) \times C^\infty_G (P, T_eG)$ restricts to ${\cal P}^\infty_G(P,\vartheta):=(\pi^*\times \nu^*)(\Gamma_{G,\vartheta}^\infty(TP))$ making it a Lie $C^\infty (M, \mathbb{R})$-submodule of $\Gamma^\infty (TM) \times C^\infty_G (P, T_eG)$. It follows from (\ref{ioilie54}) that $(X,F)\in {\cal P}^\infty_G(P,\vartheta)$ if and only if
\begin{equation}\label{lie54}
    H^*(X)\llcorner \Omega+\textbf{D}F=0.
\end{equation}
Let us note here that the condition (\ref{lie54}) is invariant with respect to the Lie $C^\infty (M, \mathbb{R})$-module operation. Thus, for $(X_1, F_1), (X_2, F_2) \in {\cal P}^\infty(P,\vartheta)$ the Lie bracket (\ref{ioi423spr}) simplifies to the form
\begin{multline}\label{ytyioi423spr}
\{(X_1,F_1),(X_2,F_2)\}_\vartheta=
([X_1,X_2], 2\Omega(H^*(X_1),H^*(X_2))-[F_1, F_2]) =\\
=([X_1,X_2], H^*(X_1)\llcorner \textbf{D}F_2-[F_1,F_2])=([X_1,X_2], H^*(X_1)(F_2)-[F_1,F_2]) .
\end{multline}

In the case when the curvature form $\Omega$ is a non-singular 2-form, i.e. when $\xi^h\llcorner\Omega=0$ implies $\xi^h=0$, one has from (\ref{lie54}) that for $(X,F)\in {\cal P}^\infty_G(P,\vartheta)$ the vector field $X \in \Gamma^\infty (TM)$ is defined uniquely by the function  $F\in C^\infty_G(P,T_eG)$. So, in this case we have the $C^\infty(M, \mathbb{R})$-modules morphism $\mathfrak{b}: C^\infty_G(P,T_eG)   \rightarrow {\cal P}^\infty_G(P,\vartheta)$. Note here that $F\in \mbox{ker}\mathfrak{b}$ if and only if $\textbf{D}F =0$, which does not mean in general that $F= \mbox{const}$. Substituting $X_1 = \mathfrak{b}(F_1)$ and $X_2 = \mathfrak{b}(F_2)$ into (\ref{ytyioi423spr}) we obtain the Lie bracket
\begin{equation}\label{221}
\{F_1, F_2 \}_\vartheta := \{(\mathfrak{b}(F_1),F_1),(\mathfrak{b}(F_2),F_2)\}_\vartheta
\end{equation}
of $F_1, F_2 \in  C^\infty_G(P,T_eG)$, which satisfies the Leibniz property
\begin{equation}\label{222}
\{F_1, fF_2 \}_\vartheta= f\{F_1, F_2 \}_\vartheta+ \mathfrak{b}(F_1)(f) F_2
\end{equation}
in sense of $C^\infty (M, \mathbb{R})$-module.

Assuming  $G=U(1)$ we find that $C^\infty_G(P,T_eG)$ is canonically isomorphic with $C^\infty(M,\mathbb R)$ and the curvature form $\Omega$ is identified with the closed $d \omega =0$ 2-form $\omega$ on $M$, which is a symplectic form in the non-singular case. Hence, formula (\ref{221}) reduces to the symplectic Poisson bracket and (\ref{222}) to its Leibniz property.

Taking the above facts into account, further we will call $({\cal P}^\infty_G(P,\vartheta), \{\cdot, \cdot \}_{\vartheta})$ the Poisson $C^\infty (M, \mathbb{R})$-module.

In the framework of the assumed terminology it is natural to consider:
\begin{enumerate}
\item[(i)] the equation (\ref{lie54}) as a generalization of Hamilton's equation to the arbitrary Lie group $G$ case;
\item[(ii)] the one-parameter group $\tau_t^{(X, F)} \in \mbox{Aut}(P,\vartheta)$ as a generalized Hamiltonian flow generated by $(X, F) \in {\cal P}^\infty_G(P,\vartheta)$ (in non-singular case by $F\in C^\infty_G(P,T_eG)$).
\end{enumerate}

In the next two sections we propose and investigate a method of quantization of the Hamiltonian flow $\tau_t^{(X, F)} \in \mbox{Aut}(P,\vartheta)$ based on the notion of $G$-equivariant positive kernel on $P\times P$.

 Though our considerations below are valid for an arbitrary Lie group $G$  we will assume that $G\subset GL(V,\mathbb{C})\cong GL(N,\mathbb{C})$ is a Lie subgroup of the linear group $GL(V,\mathbb{C})$ of a complex $N$-dimensional Hilbert space $V$. By $\langle\cdot,\cdot\rangle:V\times V\rightarrow\mathbb{C}$ we denote the scalar product for $V$ and by
 \begin{equation}\label{canac2}
    G\times V\ni (g,v)\mapsto gv\in V\nonumber
 \end{equation}
 the canonical action of $G$ in $V$. The group of unitary maps of $V$ as usually will be denoted by $U(V)\cong U(N)$.

\section{Kirillov-Kostant-Souriau prequantization morphism}

In this section we generalize the Kirillov-Kostant-Souriau prequantization procedure \cite{Ki}, \cite{Ko}, \cite{S} for the case of an arbitrary Lie group $G\subset GL(V, \mathbb{C})$, i.e. we obtain the Lie $C^\infty(M,\mathbb{R})$-module morphism $Q: {\cal P}^\infty_G(P,\vartheta) \rightarrow {\cal D}^1\Gamma^\infty(M,\mathbb{V})$ of the Poisson module ${\cal P}^\infty_G(P,\vartheta)$ into the Lie module of differential operators of the order less or equal one acting on the smooth sections of some complex vector bundle $\mathbb{V}\rightarrow M$ over $M$.

To this end we define the smooth complex bundle $\mathbb{V}:= (P\times V)/G\rightarrow M$ over $M$ associated to $P(G,\pi,M)$ by the action
$P\times V\times G\ni (p,v,g)\mapsto (pg,g^{-1}v)\in P\times V$ of the Lie group $G$ on $P\times V$. One has the natural $C^\infty(M,\mathbb{R})$-module isomorphism between the module $\Gamma^\infty(M,\mathbb{V})$ of smooth sections of $\mathbb{V}\rightarrow M$ and the module $C^\infty_G(P,V):=\{f\in C^\infty(P,V): f(pg)=g^{-1}f(p)\;for\;g\in G\}$ of $G$-equivariant smooth  functions on $P$ defined by the one-to-one dependence
\begin{equation}\label{ioiog43sec}
    \psi(\pi(p))=[(p,f(p))]:=\{(pg,g^{-1}f(p)) : g\in G  \}
\end{equation}
between $\psi\in \Gamma^\infty(M,\mathbb{V})$ and $f\in C^\infty_G(P,V)$. Using (\ref{ioiog43sec}) we define
\begin{equation}\label{ioiog43gt}
    (Q_{(X,F)}\psi)(\pi(p)):=[(p,\xi(f)(p))],
\end{equation}
for any $(X,F)=(\pi^*\times \nu^*)(\xi)\in {\cal P}^\infty_G(P,\vartheta)$ the differential operator $Q_{(X,F)}:\Gamma^\infty(M,\mathbb{V})\rightarrow\Gamma^\infty(M,\mathbb{V})$ of order less or equal one. Let us note here that if $\xi\in  \Gamma_{G,\vartheta}^\infty(TP)$ and $f\in C^\infty_G(P,V)$ then  $\xi(f)\in C^\infty_G(P,V)$.

Since $\pi^*\times \nu^*: \Gamma_{G,\vartheta}^\infty(TP)\stackrel{\sim}{\longrightarrow}{\cal P}^\infty_G(P,\vartheta)$ is an isomorphism of the Lie $C^\infty(M,\mathbb{R})$-modules, from (\ref{ioiog43sec}) and (\ref{ioiog43gt}) one obtains
\begin{equation}\label{ioipre32q1}
    [Q_{(X_{1},F_{1})},Q_{(X_{2},F_{2})}]=Q_{\{(X_{1},F_{1}),(X_{2},F_{2})\}_\vartheta},
\end{equation}
\begin{equation}\label{fffftfr32q1}
    Q_{f(X,F)}=fQ_{(X,F)},
\end{equation}
where on the left hand side of the equality (\ref{ioipre32q1}) we have the commutator of the differential operators and $\{\cdot,\cdot\}_\vartheta$ is the Lie bracket defined in (\ref{ytyioi423spr}). Since in the case  $\dim_{\mathbb{C}}V=1$ and $G=U(1)$ the Lie $C^\infty(M,\mathbb{R})$-modules monomorphism \begin{equation}\label{ddddy65y}
Q:{\cal P}^\infty(P,\vartheta)\rightarrow {\cal D}^1\Gamma^\infty(M,\mathbb{V})
\end{equation}
is the Kirillov-Kostant-Souriau prequantization morphism  we  will further extend this terminology to the general case. By ${\cal D}^1\Gamma^\infty(M,\mathbb{V})$ in (\ref{ddddy65y}) we denote the Lie $C^\infty(M,\mathbb{R})$-module of differential operators of order less or equal one acting on $\Gamma^\infty(M,\mathbb{V})$.

In order to establish a more explicit expression for $Q_{(X,F)}$, where $(X,F)\in {\cal P}^\infty(P,\vartheta)$, we will use the decomposition $\xi=\xi^h+\xi^v$ of $\xi\in \Gamma_{G,\vartheta}^\infty(TP)$ on the horizontal and vertical components. The flows $\tau_t$ and $\tau^h_t$ tangent to $\xi$ and $\xi^h$ satisfy
\begin{equation}\label{iouj87}
    \tau_t(p)=\tau^h_t(p) g(t,p),
\end{equation}
where  $g:\mathbb R\times P\rightarrow G$ is the cocycle related to the vertical (tangent to $\xi^v$) flow $\tau^v_t$ by $\tau^v_t(p)=pg(t,p)$.
\begin{proposition}\label{618191gyu}
The cocycle $g:\mathbb R\times P\rightarrow G$ corresponding to   $(X,F)\in {\cal P}^\infty_G(P,\vartheta)$ by (\ref{iouj87}) has the following form
\begin{equation}\label{znik879}
    g(t,p)=e^{tF(p)}.
\end{equation}
\end{proposition}{\it \underline{Proof}}

Since for $\xi=(\pi^*\times \nu^*)^{-1}((X, F))\in \Gamma_{G,\vartheta}^\infty(TP)$ one has  $\xi(F)=\xi(\langle \vartheta,\xi\rangle)=\langle L_\xi\vartheta,\xi\rangle+\langle \vartheta,[\xi,\xi]\rangle=0$ we find that
\begin{equation}\label{Lkop89uy6z}
    F(\tau_t(p))=F(p).
\end{equation}
Combining (\ref{Lkop89uy6z}) with (\ref{iouj87}) we obtain
\begin{equation}\label{LLKP089uy6z}
    F(\tau^h_t(p))=g(t,p)F(p)g(t,p)^{-1}.\nonumber
\end{equation}
Applying $\tau_s$ to both sides of (\ref{iouj87}) we obtain
\begin{equation}\label{HHPlj87}
    \tau_{s+t}(p)=\tau_s(\tau^h_t(p)) g(t,p)=\tau^h_s(\tau^h_t(p)) g(s,\tau^h_t(p)) g(t,p)=\tau^h_{s+t}(p) g(s,\tau^h_t(p)) g(t,p).\nonumber
\end{equation}
From the above and from (\ref{iouj87}) we find that
\begin{equation}\label{KPs43d87}
    g(s+t,p)=g(s,\tau^h_t(p)) g(t,p).\nonumber
\end{equation}
Differentiating the equality above  with respect to the parameter $s$ at $s=0$ we obtain differential equation
\begin{equation}\label{AwAwAwe43}
    \frac{d}{dt}g(t,p)=F(\tau^h_t(p))g(t,p)=g(t,p)F(p)
\end{equation}
on the cocycle $g:\mathbb R\times P\rightarrow G\subset GL(N,\mathbb C)$. The equality (\ref{znik879}) is obtained as a solution of (\ref{AwAwAwe43}) with the initial condition $g(0,p)=1\!\!1_V$.

\hfill $\Box$

For $f\in C^\infty_G(P,V)$ one has $\xi^h(f),\xi^v(f)\in C^\infty_G(P,V)$ and
\begin{equation}\label{iok90o}
    \xi^h(f)(p)=\langle df,\xi^h\rangle(p)=\langle \textbf{D}f,\xi\rangle(p),
\end{equation}
\begin{equation}\label{iok9pol}
    \xi^v(f)(p)=\langle df,\xi^v\rangle(p)=\frac{d}{dt}f(pe^{tF(p)})|_{t=0}
   =\frac{d}{dt}e^{-tF(p)}f(p)|_{t=0}=-F(p)f(p),
\end{equation}
where $\textbf{D}:C^\infty_G(P,V)\rightarrow \Gamma_G^\infty(P,T^*P\otimes V)$ is the covariant derivative of $f$ defined by
\begin{equation}\label{iom675n}
    \textbf{D}f:=df\circ pr^{hor}=df+\vartheta f.\nonumber
\end{equation}
By $pr^{hor}:TP\rightarrow T^hP$ we denoted the projection of $TP$ on its horizontal component $T^hP$.

Defining the covariant derivative $\nabla:\Gamma^\infty(M,\mathbb{V})\rightarrow\Gamma^\infty(M,T^*M\otimes\mathbb{V})$ as usual by
\begin{equation}\label{ioioipp}
    \nabla\psi(\pi(p)):=[(p,\textbf{D} f(p))]=[(p,df(p)+\vartheta(p) f(p))],
\end{equation}
where $\psi$ is defined in (\ref{ioiog43sec}), and using equations (\ref{iok90o}) and (\ref{iok9pol}) one obtains from (\ref{ioiog43gt}) the following expression
\begin{equation}\label{ioosi4e}
Q_{(X,F)}=\nabla_X-F,
\end{equation}
for the Kirillov-Kostant-Souriau  operator $Q_{(X,F)}:\Gamma^\infty(M,\mathbb{V})\rightarrow\Gamma^\infty(M,\mathbb{V})$. Let us note here that the 0-order differential operator $F$ acts on $\psi\in \Gamma^\infty(M,\mathbb{V})$ as follows
\begin{equation}\label{poioipp}
    (F\psi)(\pi(p)):=[(p,F(p) f(p))].\nonumber
\end{equation}
We note also that $F(pg) f(pg)=g^{-1}F(p) f(p)$ and $(\textbf{D}_\xi f)(pg)=g^{-1}(\textbf{D}_\xi f)(p)$.

The Kirillov-Kostant-Souriau operator is the generator
\begin{equation}\label{ioosi4ghe}
Q_{(X,F)}\psi(m):=\lim_{t\rightarrow0}\frac{1}{t}\left[(\Sigma_t\psi)(m)-\psi(m)\right]\nonumber
\end{equation}
of the one-parameter group
$\Sigma_t:\Gamma^\infty(M,\mathbb{V})\rightarrow
\Gamma^\infty(M,\mathbb{V})$ acting on the sections $\psi\in
\Gamma^\infty(M,\mathbb{V})$ by
\begin{equation}
(\Sigma_t\psi)(m):=\tau^\mathbb V_t
\psi(\sigma_{-t}(m)),\nonumber
\end{equation}
where the flows $\tau^\mathbb V_t:\mathbb{V}\rightarrow\mathbb{V}$ and $\sigma_t:M\rightarrow M$ are defined by
\begin{equation}\label{09wi8we}
    \tau^\mathbb V_t[(p,v)]:=[(\tau_t(p),v)]\nonumber
\end{equation}
and by
\begin{equation}\label{iosi89m}
    \sigma_t(\pi(p)):=\pi(\tau_t(p)),\nonumber
\end{equation}
respectively. The vector field $X\in \Gamma^\infty(TM)$  in (\ref{ioosi4e}) is tangent to the flow $\{\sigma_t\}_{t\in\mathbb R}$.

If the curvature form $\Omega$ is non-singular the  $C^\infty(M,\mathbb{R})$-module morphism $\mathfrak b:C^\infty_G(P, T_eG)\rightarrow \Gamma^\infty(TM)$
leads to the Kirillov-Kostant-Souriau prequantization morphism
\begin{equation}\label{iore650}
    Q: C^\infty_G(P,T_eG)\ni F\mapsto Q_{F}=\nabla_{X_F}-F\in {\cal D}^1\Gamma^\infty(M,\mathbb{V})\nonumber
\end{equation}
for the Poisson $C^\infty(M,\mathbb{R})$-module $(C^\infty_G(P,T_eG),\{.,.\}_\vartheta)$, where $X_F=\flat(F)$ and Poisson bracket $\{F_1,F_2\}_\vartheta$ of $F_1,F_2\in C^\infty_G(P,T_eG)$ is defined in (\ref{221}).

Furthermore we will need the vector bundle $\bar{\mathbb{V}}:= (P\times V)/G\rightarrow M$ associated to $P(G,\pi,M)$ by the action
\begin{equation}\label{ppp7s43}
    P\times V\times G\ni (p,v,g)\mapsto (pg,g^{\dagger}v)\in P\times V\nonumber
\end{equation}
as well as  the $C^\infty(M,\mathbb{C})$-module
\begin{equation}\label{de4434q}
    C^\infty_{\bar{G}}(P,V):=\{\overline{f}\in C^\infty(P,V): \overline{f}(pg)=g^{\dagger}\overline{f}(p)\;for\;g\in G\}
\end{equation}
of the $V$-valued smooth function on $P$. Similarly as in (\ref{ioiog43sec}) the equality $\bar \psi(\pi(p)):=[(p,\overline{f}(p))]$ defines the isomorphism $\Gamma^\infty(M,\bar{\mathbb{V}})\cong C^\infty_{\bar G}(P,V)$ of $C^\infty(M,\mathbb{C})$-modules. Using this isomorphism we define
\begin{equation}\label{ioiog8t}
    (\bar Q_{(X,F)}\bar \psi)(\pi(p)):=[(p,\xi(\overline{f})(p))]\nonumber
\end{equation}
another Kirillov-Kostant-Souriau differential operator acting now on $\Gamma^\infty(M,\bar{\mathbb{V}})$. For vector field $\xi\in \Gamma_{G}^\infty(TP)$ tangent to $\tau_t$ from (\ref{iouj87}) we find
\begin{equation}\label{sam777}
\xi(\overline{f})(p)=\frac{d}{dt}\overline{f}(\tau_t(p))|_{t=0}=\frac{d}{dt}\overline{f}(\tau^h_t(p))|_{t=0} +\frac{d}{dt}g(t,p)^\dagger |_{t=0}\overline{f}(p)=\bar{\textbf{D}}\overline{f}(p)+ F(p)^\dagger \overline{f}(p),\nonumber
\end{equation}
where
\begin{equation}\label{iom999n}
    \bar{\textbf{D}}\overline{f}:=d\overline{f}+\vartheta^\dagger\overline{f}.\nonumber
\end{equation}

Next, using the covariant derivative $\bar \nabla:\Gamma^\infty(M,\bar{\mathbb{V}})\rightarrow\Gamma^\infty(M,T^*M\otimes\bar{\mathbb{V}})$ defined by $\bar \nabla\bar \psi(\pi(p)):=[(p,\bar{\textbf{D}}\overline{f}(p))]$ we obtain
\begin{equation}\label{ikksi4e}
\bar Q_{(X,F)}=\bar \nabla_X+F^\dagger,
\end{equation}
the generator of the flow
\begin{equation}\label{09je43je}
(\bar \Sigma_t\bar \psi)(m):=\tau^{\bar{\mathbb{V}}}_t
\bar \psi(\sigma_{-t}(m)),
\end{equation}
where $F^\dagger\bar \psi$ and $\tau^{\bar{\mathbb{V}}}_t:\bar{\mathbb{V}}\rightarrow\bar{\mathbb{V}}$ are defined as follows

\begin{equation}\label{hhoipp}
    (F^\dagger\bar \psi)(\pi(p)):=[(p,F(p)^\dagger\overline{f}(p))]\nonumber
\end{equation}
and
\begin{equation}\label{666we}
    \tau^{\bar{\mathbb{V}}}_t[(p,v)]:=[(\tau_t(p),v)]\nonumber
\end{equation}
for $[(p,v)]\in \bar{\mathbb{V}}$.

The covariance properties (\ref{ioipre32q1}) and (\ref{fffftfr32q1}) for $\bar Q$ are proved in an analogous way as for $Q$.

 Now let us mention that using the scalar product $\langle\cdot,\cdot\rangle$ in $V$ one defines for $\bar \psi_1\in\Gamma^\infty(M,\bar{\mathbb{V}}) $ and $\psi_2\in \Gamma^\infty(M,\mathbb{V})$ the smooth function
\begin{equation}\label{llll}
\langle\!\langle\bar \psi_1,\psi_2 \rangle\!\rangle(\pi(p)):=\langle \overline{f}_1(p),f_2(p)\rangle
\end{equation}
on $M$. One sees from $\langle\!\langle\bar \psi_1,\psi_2 \rangle\!\rangle(\sigma_t(\pi(p)))=\langle \overline{f}_1(\tau_t(p)),f_2(\tau_t(p))\rangle$ that the following property
\begin{equation}\label{pp7pp}
X(\langle\!\langle\bar \psi_1,\psi_2 \rangle\!\rangle)=\langle\!\langle\bar Q_{(X,F)}\bar \psi_1,\psi_2 \rangle\!\rangle+\langle\!\langle\bar \psi_1,Q_{(X,F)}\psi_2 \rangle\!\rangle\nonumber
\end{equation}
for the pairing (\ref{llll}) is valid.

Fixing a local trivialization $s_\alpha:{\cal O}_\alpha\rightarrow P$ (where $\bigcup_{\alpha\in I} {\cal O}_\alpha=M$ is an open covering of $M$) of the principal bundle $P(G,\pi,M)$ one defines the local cocycles $g_\alpha(t,\cdot):{\cal O}_\alpha\rightarrow G$ and $h_\alpha(t,\cdot):{\cal O}_\alpha\rightarrow G$ by
\begin{equation}\label{ioi1coc324}
 \tau_t(s_\alpha(m))=s_\alpha(\sigma_t(m))g_\alpha(t,m),
\end{equation}
\begin{equation}\label{oio1coc324}
 \tau_t^h(s_\alpha(m))=s_\alpha(\sigma_t(m))h_\alpha(t,m)
\end{equation}
for sufficiently small $t$. From (\ref{iouj87}) and (\ref{znik879}) one has
\begin{equation}\label{iuiouj87}
    \tau_t(s_\alpha(m))=\tau^h_t(s_\alpha(m)) e^{tF(s_\alpha(m))}.
\end{equation}
Hence, from (\ref{ioi1coc324})-(\ref{iuiouj87}) one obtains
\begin{equation}\label{7u7u7u}
g_\alpha(t,m)=h_\alpha(t,m)\exp(tF(s_\alpha(m))).
\end{equation}
From (\ref{7u7u7u}) and the cocycle properties
\begin{equation}\label{7u7u7uoio}
g_\alpha(t+s,m)=g_\alpha(s,\sigma_t(m))g_\alpha(t,m),\nonumber
\end{equation}
\begin{equation}
h_\alpha(t+s,m)=h_\alpha(s,\sigma_t(m))h_\alpha(t,m)\nonumber
\end{equation}
we have
\begin{equation}\label{io7u7u7uioi}
h_\alpha(t,m)e^{sF(s_\alpha(m))}=e^{sF(s_\alpha(\sigma_t(m)))}h_\alpha(t,m),\nonumber
\end{equation}
which is equivalent to
\begin{equation}\label{io7u7u7u}
h_\alpha(t,m)F(s_\alpha(m))=F(s_\alpha(\sigma_t(m)))h_\alpha(t,m).\nonumber
\end{equation}
In order to obtain $\frac{d}{dt}h_\alpha(t,m)|_{t=0}\in T_eG$ we note that from (\ref{oio1coc324}) it  follows that
\begin{equation}\label{fffri}
    \xi^h(s_\alpha(m))=Ts_\alpha(m)X(m)+T\kappa_{s_\alpha(m)}(e)\frac{d}{dt}h_\alpha(t,m)|_{t=0}.
\end{equation}
Next, applying $\vartheta(s_\alpha(m))$ to both sides of (\ref{fffri}) and using (\ref{kon4a}) we obtain
\begin{equation}\label{rerety7}
    \frac{d}{dt}h_\alpha(t,m)|_{t=0}=-\vartheta(s_\alpha(m))(Ts_\alpha(m)X(m))    =- \vartheta_\alpha(m)( X(m)),\nonumber
\end{equation}
where $\vartheta_\alpha(m):=\vartheta(s_\alpha(m))\circ Ts_\alpha(m)=(s_\alpha^*\vartheta)(m)$.

Introducing the notations $\phi_\alpha(m):=\frac{d}{dt}g_\alpha(t,m)|_{t=0}$ and $F_\alpha(m):=F\circ s_\alpha(m)$, after differentiating (\ref{7u7u7u}) at $t=0$ we obtain the equality
\begin{equation}\label{ftery656}
\phi_\alpha(m)=-\langle \vartheta_\alpha,X\rangle(m)+F_\alpha(m),
\end{equation}
which is useful for finding  the local expression for the infinitesimal generator of the flow $\Sigma_t:\Gamma^\infty(M,\mathbb{V})\rightarrow
\Gamma^\infty(M,\mathbb{V})$ defined in (\ref{09je43je}). Namely, we have
\begin{equation}\label{sig543}
(\Sigma_t\psi)(m)=\tau^\mathbb V_t
\psi(\sigma_{-t}(m))=\tau^\mathbb V_t[s_\alpha(\sigma_{-t}(m)),
(f\circ s_\alpha)(\sigma_{-t}(m)) ]=\nonumber
\end{equation}
\begin{equation}\label{sig544}
=[\tau_ts_\alpha(\sigma_{-t}(m)), (f\circ
s_\alpha)(\sigma_{-t}(m)) ]=
[s_\alpha(m)g_\alpha(t,\sigma_{-t}(m)), (f\circ
s_\alpha)(\sigma_{-t}(m)) ]=\nonumber
\end{equation}
\begin{equation}\label{sig546}
=[s_\alpha(m),g_\alpha(t,\sigma_{-t}(m)) (f\circ
s_\alpha)(\sigma_{-t}(m)) ].
\end{equation}
Thus for $f_\alpha:=f\circ s_\alpha:{\cal O}_\alpha\rightarrow V$ one has
\begin{equation}\label{lok98ci}
(\Sigma_t f_\alpha)(m)=g_\alpha(-t,m)^{-1}
f_\alpha(\sigma_{-t}(m)).
\end{equation}

Differentiating both sides of (\ref{lok98ci})  at $t=0$  and using (\ref{ftery656}) we obtain the local representation
\begin{equation}\label{kisi4el}
(Q_{(X,F)}f_\alpha)(m)=-X(f_\alpha)(m)+\phi_\alpha(m)f_\alpha(m)=\nonumber
\end{equation}
\begin{equation}\label{kisi4eft}
=-(X+\langle \vartheta_\alpha,X\rangle)(f_\alpha)(m)+F_\alpha(m) f_\alpha(m)=\nonumber
\end{equation}
\begin{equation}\label{kisi4e}
=-(\nabla_X^\alpha f_\alpha)(m)+F_\alpha(m) f_\alpha(m),
\end{equation}
of the Kirillov-Kostant-Souriau prequantization operator $Q_{(X,F)}:\Gamma^\infty(M,\mathbb{V})\rightarrow\Gamma^\infty(M,\mathbb{V})$, where $\nabla_X^\alpha:=X+\langle \vartheta_\alpha,X\rangle$ is the local form of the covariant derivative $\nabla$ defined in (\ref{ioioipp}). Similarly we have
\begin{equation}\label{kibudel}
(\bar Q_{(X,F)}\overline{f}_\alpha)(m)=-(\nabla_X^\alpha \overline{f}_\alpha)(m)+\overline{f}_\alpha(m)F_\alpha(m)^\dag
\end{equation}
for $\bar Q_{(X,F)}:\Gamma^\infty(M,\bar{\mathbb{V}})\rightarrow\Gamma^\infty(M,\bar{\mathbb{V}})$.

In the local gauge
the Hamilton equation (\ref{lie54}) assumes the form
\begin{equation}\label{hjam8i}
    X\llcorner\Omega_\alpha+\textbf{D}F_\alpha=0,
\end{equation}
where
\begin{equation}\label{pre34d}
\Omega_\alpha:=(s_\alpha)^*\Omega=
d\vartheta_\alpha+\frac{1}{2}[\vartheta_\alpha,\vartheta_\alpha],
\end{equation}
\begin{equation}\label{pre34df}
\textbf{D}F_\alpha:= dF_\alpha+[\vartheta_\alpha,F_\alpha].
\end{equation}

Ending this subsection let us mention the well known equivariance formulae with respect to the gauge transformation
\begin{equation}\label{igtu765h}
    s_\beta(m)=s_\alpha(m)g_{\alpha\beta}(m),
\end{equation}
where $g_{\alpha\beta}:{\cal O}_\alpha\cap{\cal O}_\beta\rightarrow G$ is the respective transition cocycle, i.e. $g_{\alpha\beta}(m)g_{\beta\gamma}(m)=g_{\alpha\gamma}(m)$.
Namely, one has
\begin{equation}\label{ioiju87u}
    \vartheta_\beta(m)=g^{-1}_{\alpha\beta}(m)\vartheta_\alpha(m) g_{\alpha\beta}(m)+g^{-1}_{\alpha\beta}(m)(dg_{\alpha\beta})(m),
\end{equation}
\begin{equation}\label{ioiju87u5}
    F_\beta(m)=g^{-1}_{\alpha\beta}(m)F_\alpha(m) g_{\alpha\beta}(m),
\end{equation}
\begin{equation}\label{ioiju47u5u}
    \Omega_\beta(m)=g^{-1}_{\alpha\beta}(m)\Omega_\alpha(m) g_{\alpha\beta}(m),
\end{equation}
\begin{equation}\label{ioiju97u9}
    f_\beta(m)=g^{-1}_{\alpha\beta}(m)f_\alpha(m) ,
\end{equation}
\begin{equation}\label{ioiju80u}
    \phi_\beta(m)=g^{-1}_{\alpha\beta}(m)\phi_\alpha(m) g_{\alpha\beta}(m)-g^{-1}_{\alpha\beta}(m)(Xg_{\alpha\beta})(m),
\end{equation}
where $m\in{\cal O}_\alpha\cap{\cal O}_\beta$.

\section{Positive definite kernels and quantization}\label{section4}

In the previous section we obtained the formulas (\ref{ioosi4e}) and (\ref{ikksi4e}) on Kirillov-Kostant-Souriau prequantization operators $Q_{(X,F)}$ and $\bar Q_{(X,F)}$ as well as their local versions (\ref{kisi4e}) and (\ref{kibudel}). Here we will present a procedure which allows us to treat them as  self-adjoint operators in a Hilbert space. This quantization procedure is based  on the notion of $G$-equivariant positive kernel, see \cite{O-H}. In order to make the paper self-sufficient we present the above procedure in detail. Some new results complementary to the ones in \cite{O-H} will also be presented in this section.

The theory of reproducing kernels for like-Hermitian smooth vector bundles over smooth Banach manifolds and related linear connections one can find in \cite{B-G1}, \cite{B-G2}.

Let us recall here that we have assumed that $\dim_\mathbb{C} V = N <+\infty$ and $G \subset GL(V)\cong GL(N, \mathbb{C})$. Further, by $\mathcal{B}(V) \cong \mbox{Mat}_{N \times N} (\mathbb{C})$ we denote the $C^*$-algebra of linear maps of $V$ and by $\mathcal{B}(V, \mathcal{H})$ the right Hilbert $\mathcal{B}(V)$-module of linear maps $\Gamma: V \rightarrow \mathcal{H}$ from the Hilbert space $V$ into the separable Hilbert space $\mathcal{H}$. Let us note here that from $\dim_\mathbb{C} V<+\infty$ follows boundness of $\Gamma:V\rightarrow\mathcal{H}$. The $\mathcal{B}(V)$-valued scalar product $\langle \cdot ; \cdot \rangle : \mathcal{B}(V, \mathcal{H})\times \mathcal{B}(V, \mathcal{H}) \rightarrow \mathcal{B}(V)$ on $\mathcal{B}(V, \mathcal{H})$ is defined by
\begin{equation}
\langle \Gamma ; \Delta \rangle := \Gamma^*\Delta\nonumber
\end{equation}
for $\Gamma, \Delta \in \mathcal{B}(V, \mathcal{H})$, where the operator $\Gamma^*: \mathcal{H}\rightarrow V$ is the one adjoint to $\Gamma:V\rightarrow \mathcal{H}$.

\begin{definition}\label{def41}
A smooth map $\mathfrak{K}:P\rightarrow \mathcal{B}(V,{\cal H})$ will be called a $G$-equivariant coherent state map if it has the following properties:
\begin{enumerate}
\item[(i)] the $G$-equivariance property, i.e.
\begin{equation}\label{ifqdy43e}
\mathfrak{K}(pg)=\mathfrak{K}(p)g
\end{equation}
for any $p\in P$ and for any $g\in G$;
\item[(ii)] non-singularity, i.e.
\begin{equation}\label{qa123b3}
\ker \mathfrak{K}(p)=\{0\},\;\;or\;\;equivalently\;\;\mathfrak{K}(p)^*\mathfrak{K}(p)\in GL(V,\mathbb{C})
\end{equation}
for any $p\in P$;
\item[(iii)] the set $\mathfrak{K}(P)V$ is linearly dense in $\cal H$, i.e.
\begin{equation}\label{ffke5r}
    \bigcap_{p\in P}\ker \mathfrak{K}(p)^*=\{0\},\;\;or\;\;equivalently\;\;\{\mathfrak{K}(p)v: p\in P \;\;and\;\; v\in V \}^\perp=\{0\}.
\end{equation}
\end{enumerate}
\end{definition}
By $\mathfrak{K}(p)^*$ in (\ref{qa123b3}) and (\ref{ffke5r}) we denoted the map $\mathfrak{K}(p)^*: {\cal H}\rightarrow V$ adjoint to $\mathfrak{K}(p):V\rightarrow {\cal H}$, i.e. such that $\langle\psi|\mathfrak{K}(p)v\rangle=\langle\mathfrak{K}(p)^*\psi,v\rangle$ for $\psi\in{\cal H}$ and $v\in V$, where $\langle\cdot|\cdot\rangle$ is the scalar product in Hilbert space ${\cal H}$. For the existence of $\mathfrak{K}:P \rightarrow B(V, \mathcal{H})$ with the above properties see \cite{N-R}.
\begin{definition}\label{jadro}
A smooth map $K:P\times P\rightarrow{\cal B}(V)$ will be called a $G$-equivariant positive definite kernel if it has the following properties
\begin{enumerate}
\item[(i)] the $G$-equivariance property, i.e.
\begin{equation}\label{qaf21b3}
K(p,qg)= K(p,q)g
\end{equation}
for any $p,q\in P$ and $g\in G$;
\item[(ii)] non-singularity, i.e.
\begin{equation}\label{qa12f3b3}
K(p,p)\in GL(V,\mathbb{C})\subset \mathcal{B}(V)
\end{equation}
for any $p\in P$;
\item[(iii)]  positivity, i.e.
\begin{equation}\label{qb5fiu}
\sum_{i,j=1}^J  \langle v_i, K(p_i,p_j)v_j\rangle\geqq 0
\end{equation}
for arbitrary finite sequences $p_1,\ldots,p_J\in P$ and $v_1,\ldots,v_J\in V$.
\end{enumerate}
\end{definition}
From the positivity condition (\ref{qb5fiu}) one obtains the following properties of the kernel
\begin{equation}\label{qfele}
    K(p,q)^\dagger=K(q,p)\nonumber
\end{equation}
and
\begin{equation}\label{qpa21b3}
K(ph,qg)=h^{\dagger} K(p,q)g,\nonumber
\end{equation}
where $g,h\in G$ and $p,q\in P$.

The above structures are mutually dependent. Extending the rather well known scheme from the theory of reproducing (positive) kernels  \cite{N}, \cite{Ar} to this more geometrically complicated setting we shortly describe this dependence.

Starting from the coherent state map $\mathfrak{K}:P\rightarrow \mathcal{B}(V,{\cal H})$ we define the $G$-equivariant positive definite kernel by
\begin{equation}\label{f6qf4r54}
    K(p,q):=\mathfrak{K}(p)^*\mathfrak{K}(q).
\end{equation}
The smoothness of the kernel (\ref{f6qf4r54}) and the properties (\ref{qaf21b3})-(\ref{qb5fiu}) follow from the smoothness of $\mathfrak{K}:P\rightarrow \mathcal{B}(V,{\cal H})$ and its properties mentioned in (\ref{ifqdy43e})-(\ref{ffke5r}).

The opposite dependence needs longer considerations. Firstly let us define the vector space
\begin{equation}\label{Afr432}
    {\cal D}_K:=\left\{\bar f=\sum_{i\in {J}} K(\cdot,q_i)v_i:\;q_i\in P,\;v_i\in V\right\}
\end{equation}
 of $V$-valued functions on $P$, where $J$ is a finite set of indices, i.e. the vector subspace ${\cal D}_K\subset C^\infty_{\bar G}(P,V)$ which consists of linear combinations of the functions $K(\cdot,q)v$ indexed by $q\in P$ and $v\in V$. In order to define the scalar product $ \langle \bar f_1|\bar f_2\rangle_{K}$ of $\bar f_1,\bar f_2\in {\cal D}_K$ we extend the summations in $\bar f_1=\sum_{i\in J_1}K(\cdot,q^1_i)v^1_i$ and $\bar f_2=\sum_{j\in J_2}K(\cdot,q^2_j)v^2_j$ to the set of indexes  $J=J_1\cup J_2$ and define  $\{p_1,\ldots,p_J \}:=\{q^1,\ldots,q^1_{J_1} \}\cup\{q^2_1,\ldots,q^2_{J_2} \}$. After that  assuming $v^1_j=0$ for $j\in J\setminus J_1$ and $v^2_j=0$ for $j\in J\setminus J_2$ we define the scalar product
\begin{equation}\label{sc3e}
    \langle \bar f_1|\bar f_2\rangle_{K}:=\sum_{i,j\in J}\langle v_i^1,
    K(p_i,p_j)v^2_j\rangle,
\end{equation}
of $\bar f_k=\sum_{j\in J}K(\cdot,p_j)v^k_j$, where $k=1,2$.

In particular one has
\begin{equation}\label{dddd3}
\langle \bar f|K(\cdot,p)v\rangle_{K}=\sum_{i\in J}\langle v_i,
    K(p_i,p)v\rangle=\langle \bar f(p),v\rangle.
\end{equation}
From (\ref{dddd3}) and the Schwarz inequality for the scalar product (\ref{sc3e}) one obtains
\begin{equation}\label{r6e43}
    |\langle \bar f(p),v\rangle|=|\langle \bar f| K(\cdot,p)v\rangle_K|\leq \langle \bar f|\bar f\rangle_K^{1/2}\langle K(\cdot,p)v| K(\cdot,p)v\rangle_K^{1/2}\leq\nonumber
\end{equation}
\begin{equation}\label{r6e4fg3k}
\leq \|f\|_K \langle v,K(p,p)v\rangle^{1/2}\leq \|f\|_K\| K(p,p)\|^{1/2}\|v\|.
\end{equation}
From the inequality (\ref{r6e4fg3k}) one sees that $\|f\|_K=\langle f| f\rangle_K^{1/2}=0$ implies $f=0$, so, $\|\cdot\|_K$ is a norm on ${\cal D}_K$.

We conclude from (\ref{r6e4fg3k}) that the evaluation functional $E_p(\bar f):=\bar f(p)$ satisfies
\begin{equation}\label{f4f43q}
    \|E_p(\bar f)\|\leq\|K(p,p)\|^{1/2}\|\bar f\|_K,\nonumber
\end{equation}
i.e. it is a bounded functional on ${\cal D}_K$ for every $p\in P$. So, we can extend it to the Hilbert space ${\cal H}_{K}\supset {\cal D}_K$ being the abstract extension of the pre-Hilbert space ${\cal D}_K\subset C^\infty_{\bar G}(P,V)$. It follows from (\ref{r6e4fg3k}) that for any equivalence class $[\{\bar f_n\}]\in {\cal H}_{K}$ of  Cauchy sequences $\{\bar f_n\}\subset {\cal D}_K$ one defines
\begin{equation}\label{je332}
    \bar f(p):=\lim_{n\rightarrow\infty}\bar f_n(p)\nonumber
\end{equation}
a function $\bar f:P\rightarrow V$ depending on $[\{\bar f_n\}]$ only. Hence we see that the Hilbert space ${\cal H}_{K}$ is realized in a natural way as a vector subspace of the vector space of $V$-valued functions on $P$ which satisfy the $G$-equivalence condition from (\ref{de4434q}).

Now, rewriting (\ref{dddd3}) as follows
\begin{equation}\label{dyygd3}
\langle \bar f|K(\cdot,p)v\rangle_{K}=\langle \bar f|E_p^*v\rangle_{K},\nonumber
\end{equation}
where $\bar f\in{\cal H}_{K}$, $v\in V$ and $E_p^*:V\rightarrow{\cal H}_{K} $ is the conjugation of $E_p:{\cal H}_{K}\rightarrow V $, we define the coherent state map $\mathfrak{K}_K:P\rightarrow \mathcal{B}(V,{\cal H}_{K})$ by
\begin{equation}\label{kr4e}
\mathfrak{K}_K(p):=E_p^*=K(\cdot,p).\nonumber
\end{equation}
One easily sees that the properties (\ref{ifqdy43e})-(\ref{ffke5r}) for $\mathfrak{K}_K:P\rightarrow \mathcal{B}(V,{\cal H}_{K})$ follow from the ones for $K$ given in (\ref{qaf21b3})-(\ref{qb5fiu}). The smoothness of $\mathfrak{K}_K:P\rightarrow \mathcal{B}(V,{\cal H}_{K})$ follows from the smoothness of the positive kernel $K$ (see Proposition 2.1 in \cite{O-H}).

As we see from (\ref{f6qf4r54}) a coherent state map $\mathfrak{K}:P\rightarrow \mathcal{B}(V,{\cal H})$ defines the positive kernel $K:P\times P\rightarrow {\cal B}(V)$. The Hilbert space ${\cal H}_{K}$ for this kernel is isomorphic  to the Hilbert space ${\cal H}$, where the isomorphism $I_K:{\cal H}\stackrel{\sim}{\longrightarrow}{\cal H}_{K}$ is defined by
\begin{equation}\label{qqqwq}
I_K:{\cal H}\ni|\psi\rangle\rightarrow I_K(|\psi\rangle):=\mathfrak{K}(\cdot)^*|\psi\rangle\in{\cal H}_{K}.
\end{equation}
In order to see that $I_K$ is indeed an isomorphism we note that for $|\psi\rangle\in {\cal D}_{\mathfrak{K}}$, where ${\cal D}_{\mathfrak{K}}\subset {\cal H}$ is defined by
\begin{equation}\label{q}
{\cal D}_{\mathfrak{K}}:=\left\{|\psi\rangle=\sum_{j\in J} \mathfrak{K}(p_j)v_j:\;p_j\in P,\;v_j\in V\right\},
\end{equation}
one has $I_K|\psi\rangle=\sum_{j\in J} K(\cdot,p_j)v_j $ and $\langle I_K|\psi\rangle|I_K|\psi\rangle\rangle_K=\langle\psi|\psi\rangle_{{\cal H}}$, i.e.  $I_K:{\cal D}_{\mathfrak{K}}\rightarrow{\cal D}_K$ is a linear isometry of dense linear subspaces ${\cal D}_{\mathfrak{K}}\subset {\cal H}$ and ${\cal D}_K\subset {\cal H}_K$, so it extends to the isomorphism of the Hilbert spaces.

The isometry $I_K:{\cal H}\rightarrow{\cal H}_{K}$ defines the isomorphism of Banach spaces $L_{I_K}:\mathcal{B}(V,{\cal H})\rightarrow\mathcal{B}(V,{\cal H}_{K})$ by $L_{I_K}A:={I_K}\circ A$, where $A\in \mathcal{B}(V,{\cal H})$.

Therefore, if the coherent state map $\mathfrak K$ and the positive definite kernel $K$ are related by (\ref{f6qf4r54}), then
\begin{equation}\label{f5f5f5}
    \mathfrak K_K=I_K\circ\mathfrak K.\nonumber
\end{equation}
If $\mathfrak K_K=I_{K_1}\circ\mathfrak K_1=I_{K_2}\circ\mathfrak K_2$, then $\mathfrak K_2=L_{{I_{K_2}^{-1}}\circ I_{K_1}}\mathfrak K_1$, i.e. the positive definite kernel $K$ defines the coherent state map $\mathfrak K$  up to an isomorphism of the Hilbert space $\mathcal{H}$.

Now let us  consider the tautological bundle $\pi_{N}:\mathbb{E}_N\rightarrow \mbox{Grass}(N,{\cal H})$ over the Grassmanian of $N$-dimensional subspaces of the Hilbert space ${\cal H}$. By definition the total space of this bundle is
\begin{equation}\label{ioid6}
    \mathbb{E}_N:=\{(\gamma,q)\in {\cal H}\times\mbox{Grass}(N,{\cal H}): \gamma\in q\}\nonumber
\end{equation}
and $\pi_{N}:= pr_2|_{\mathbb{E}_N}$, where $pr_2$ is the projection of ${\cal H}\times\mbox{Grass}(N,{\cal H})$ on its second component.

The coherent state map $\mathfrak{K}:P\rightarrow \mathcal{B}(V,{\cal H})$ defines the following morphisms
\unitlength=5mm \begin{equation}\label{dia333e3}
\begin{picture}(11,4.6)
        \put(-2,4){\makebox(0,0){$\bar{\mathbb{V}}$}}
 \put(-2,0){\makebox(0,0){$M$}}
                \put(5,4){\makebox(0,0){$\mathbb{V}$}}
       \put(5,0){\makebox(0,0){$M$}}
 \put(12,4){\makebox(0,0){$\mathbb{E}_N$}}
    \put(12,0){\makebox(0,0){$\mbox{Grass}(N,{\cal H})$}}

    \put(5,3){\vector(0,-1){2}}
    \put(-2,3){\vector(0,-1){2}}
    \put(12,3){\vector(0,-1){2}}
    \put(6,4){\vector(1,0){4.9}}
    \put(6.3,0){\vector(1,0){3}}
 \put(-0.8,0){\vector(1,0){4.5}}
 \put(-0.5,4){\vector(1,0){4}}
      \put(8.5,4.8){\makebox(0,0){${[\mathfrak{K}]}_N$}}
    \put(8,0.5){\makebox(0,0){$[\mathfrak{K}]$}}
\put(1.5,0.5){\makebox(0,0){$ id$}}
\put(1.5,4.8){\makebox(0,0){${[K^{-1}]}$}}
\put(-3,2){\makebox(0,0){$\pi_{\bar{\mathbb{V}}}$}}
\put(4,2){\makebox(0,0){$\pi_{\mathbb{V}}$}}
\put(11,2){\makebox(0,0){$\pi_{N}$}}
    \end{picture}
    \end{equation}
of vector bundles, where the morphism $[K^{-1}]:\bar{\mathbb{V}}\rightarrow\mathbb{V}$ is defined by
\begin{equation}\label{gggghj}
    \bar{\mathbb{V}}\ni[(p,v)]\longrightarrow[K^{-1}]([(p,v)]):=[(p,K(p,p)^{-1}v)]\in\mathbb{V}.
\end{equation}
The horizontal arrows on the right hand side of (\ref{dia333e3}) are defined by
\begin{equation}\label{ioid65r}
   M\ni[p]=\pi^{-1}(\pi(p))\longrightarrow [\mathfrak{K}]([p]):=\mathfrak{K}(p)V\in\mbox{Grass}(N,{\cal H})\nonumber
\end{equation}
and by
\begin{equation}\label{ioid66y}
    \mathbb{V}\ni[(p,v)]\longrightarrow[\mathfrak{K}]_N([(p,v)]):= (\mathfrak{K}(p)v,[\mathfrak{K}]([p]))\in\mathbb{E}_N.
\end{equation}
Correctness of the above definitions follows from the defining properties (\ref{ifqdy43e})-(\ref{ffke5r}) of the coherent state map $\mathfrak{K}:P\rightarrow \mathcal{B}(V,{\cal H})$. The bundle morphism defined in  (\ref{gggghj}) is an identity covering isomorphism of vector bundles with inverse given by $[K]:\mathbb{V}\rightarrow\bar{\mathbb{V}}$.

Restricting the scalar product $\langle\cdot|\cdot\rangle$ of ${\cal H}$ to the fibers $\pi_N^{-1}(q)$ of the tautological vector bundle $\pi_{N}:\mathbb{E}_N\rightarrow \mbox{Grass}(N,{\cal H})$ and using the vector bundles morphism (\ref{ioid66y}) one defines
\begin{equation}\label{ioi9j}
    H_{[p]}([(p,v)],[(p,w)]):=\langle\mathfrak{K}(p)v |\mathfrak{K}(p)w \rangle=\langle K(p,p)v, w\rangle
\end{equation}
 the Hermitian structure on the vector bundle $\pi_{\mathbb{V}}:\mathbb{V}\rightarrow M$.

\begin{proposition}\label{lkl99990}
The coherent state map $\mathfrak{K}:P\rightarrow \mathcal{B}(V,{\cal H})$ defines by
\begin{equation}\label{ioiqqon45}
    \vartheta(p)_{\mathfrak{K}}:=(\mathfrak{K}(p)^*\mathfrak{K}(p))^{-1}\mathfrak{K}(p)^*d\mathfrak{K}(p)
\end{equation}
the metric (with respect to the Hermitian structure (\ref{ioi9j}))  connection form $\vartheta_{\mathfrak{K}}\in\Gamma^\infty(T^*P\otimes T_eG)$ on the $G$-principal bundle $P(G,\pi,M)$.
\end{proposition}{\it \underline{Proof}}

From the equivariance property (\ref{ifqdy43e}) one has
\begin{equation}\label{ioi9i9}
    \vartheta_{\mathfrak{K}}(pg)=(\mathfrak{K}(pg)^*\mathfrak{K}(pg))^{-1}\mathfrak{K}(pg)^*d\mathfrak{K}(pg)
    =\nonumber
\end{equation}
\begin{equation}\label{iggoi9i9}
    =g^{-1}(\mathfrak{K}(p)^*\mathfrak{K}(p))^{-1}(g^*)^{-1}g^*\mathfrak{K}(p)^*d\mathfrak{K}(p)g
    =g^{-1}\vartheta_{\mathfrak{K}}(p)g,\nonumber
\end{equation}
for $g\in G$, so $\vartheta_{\mathfrak{K}}$ satisfies (\ref{kon4}).

In order to show the condition (\ref{kon4a}) we take $X:=\frac{d}{dt}g(t)|_{t=0}$, where $]-\varepsilon,\varepsilon[\ni t\mapsto g(t)\in G$ is a smooth curve in $G$ such that $g(0)=e$, and substitute $T\kappa_p(e)X$ into both sides of the definition (\ref{ioiqqon45}). Thus we obtain
\begin{equation}\label{ioikon4ja}
    \vartheta_{\mathfrak{K}}(p)(T\kappa_p(e)X)=
    (\mathfrak{K}(p)^*\mathfrak{K}(p))^{-1}\mathfrak{K}(p)^*d\mathfrak{K}(p)(T\kappa_p(e)X)=\nonumber
    \end{equation}
\begin{equation}
    =(\mathfrak{K}(p)^*\mathfrak{K}(p))^{-1}\mathfrak{K}(p)^*d(\mathfrak{K}\circ\kappa_p)(e)X=
    (\mathfrak{K}(p)^*\mathfrak{K}(p))^{-1}\mathfrak{K}(p)^*\frac{d}{dt}\mathfrak{K}(pg(t))|_{t=0}=\nonumber
\end{equation}
\begin{equation}\label{ioikon4a}
    =
    (\mathfrak{K}(p)^*\mathfrak{K}(p))^{-1}\mathfrak{K}(p)^*\mathfrak{K}(p)\frac{d}{dt}g(t)|_{t=0}=X.
\end{equation}
From (\ref{ioikon4a}) we see also that $T_pP=\ker \vartheta_{\mathfrak{K}}(p)\oplus T^v_p P$ and $\langle\vartheta_{\mathfrak{K}}(p),\xi(p)\rangle\in T_eG$ for $\xi(p)\in T_pP$.

It follows from
\begin{equation}\label{ioik3d}
    dK(p,p)=K(p,p)\vartheta_{\mathfrak{K}}(p) +\vartheta_{\mathfrak{K}}(p)^* K(p,p),
\end{equation}
where $K(p,p)=\mathfrak{K}(p)^*\mathfrak{K}(p)$, that $\vartheta_{\mathfrak{K}}$ is a metric connection with respect to the Hermitian structure (\ref{ioi9j}).
In order to see this let us take $f_1,f_2\in C_G^\infty(P,V)$. Then from (\ref{ioik3d}) one obtains
\begin{equation}
    d\langle f_1(p), K(p,p)f_2(p)\rangle=\nonumber
 \end{equation}
 \begin{equation} \label{ioik3ju}
    =\langle df_1(p)+\vartheta(p) f_1(p), K(p,p)f_2(p)\rangle +\langle f_1(p), K(p,p)(df_2(p)+\vartheta(p) f_2(p))\rangle.
\end{equation}
Since arbitrary sections $\psi_1,\psi_2\in\Gamma^\infty(M,\mathbb{V})$ of the vector bundle $\pi_{\mathbb{V}}:\mathbb{V}\rightarrow M$ can be written as $\psi_i(\pi(p))=[(p,f_i(p))]$, $i=1,2$, from (\ref{ioik3ju}) we obtain
\begin{equation}\label{ioip03d}
    dH(\psi_1,\psi_2)=H(\nabla \psi_1,\psi_2)+ H(\psi_1,\nabla\psi_2),\nonumber
\end{equation}
where $\nabla:\Gamma^\infty(M,\mathbb{V})\rightarrow\Gamma^\infty(M,T^*M\otimes\mathbb{V})$ is the covariant derivative defined in (\ref{ioioipp}).

\hfill $\Box$

After these preliminary considerations we propose a method of quantization of the generalized Hamiltonian flow $\tau_t^{(X,F)}:P\rightarrow P$ tangent to the vector field $\xi=(\pi^*\times \nu^*)^{-1}(X, F)\in\Gamma_{G,\vartheta}^\infty(TP)$, where $(X,F)\in {\cal P}^\infty_G(P,\vartheta)$.

\begin{definition}\label{quantization7}
A strongly continuos unitary representation $U^{(X,F)}:(\mathbb{R},+)\rightarrow(\mbox{Aut}\;{\cal H},\circ)$ of the additive group $(\mathbb{R},+)$ is the positive kernel (coherent state) quantization of a generalized Hamiltonian flow $\tau^{(X,F)}:(\mathbb{R},+)\rightarrow\mbox{Aut}(P,\vartheta)$ if there exists a coherent state map $\mathfrak{K}:P\rightarrow \mathcal{B}(V,{\cal H})$ such that:
\begin{enumerate}
\item[(i)] the connection form $\vartheta$ is equal  to $\vartheta_{\mathfrak{K}}$ defined in (\ref{ioiqqon45});
\item[(ii)] the equivariance property
\begin{equation}\label{ioig5r43}
    \mathfrak{K}(\tau_t^{(X,F)}(p))=U^{(X,F)}_t\mathfrak{K}(p)
\end{equation}
for $\mathfrak{K}:P\rightarrow \mathcal{B}(V,{\cal H})$ is fulfilled with respect to both considered flows.
\end{enumerate}
\end{definition}

From the condition (iii), see (\ref{ffke5r}) of the Definition \ref{def41} it follows that $\tau_t^{(X,F)}$ defines $U^{(X,F)}_t$ in an unique way.

Let us also note that if $\mathfrak{K}:P\rightarrow \mathcal{B}(V,{\cal H})$ is a one-to-one smooth map and the flow $\tau_t\in\mbox{Aut}(P)$ satisfies $\mathfrak{K}(\tau_t(p))=U_t\mathfrak{K}(p)$ for a certain unitary flow $U_t$, then $\tau_t\in\mbox{Aut}(P,\vartheta_{\mathfrak{K}})$ and thus there exists
$(X,F)\in \Gamma^\infty(TM)\times C^\infty_{G,\vartheta_{\mathfrak{K}}}(P,T_eG)$ such that $\tau_t=\tau_t^{(X,F)}$ and $U_t=U^{(X,F)}_t$. Additionally $(X,F)$ satisfies the generalized Hamilton equation (\ref{lie54}) for the curvature form $\Omega_{\mathfrak{K}}$ defined by $\vartheta_{\mathfrak{K}}$. If $\Omega_{\mathfrak{K}}$ is non-singular then the vector field $X\in\Gamma^\infty(TM)$ is uniquely  defined by  $F\in C^\infty_G(P,T_eG)$.

Since for the non-singular curvature form $\Omega$ the equation (\ref{lie54}) allows one to define $X\in\Gamma^\infty(TM)$ by $F\in C^\infty_G(P,T_eG)$, so, in this case we will use the notation $\tau^F_t$ and $U^F_t$ instead of $\tau^{(X,F)}_t$ and $U^{(X,F)}_t$.

It follows from Stone's Theorem, see e.g. \cite{R-S}, that there exists a self-adjoint operator $\widehat{F}$ with domain ${\cal D}_{\widehat{F}}$ dense in ${\cal H}$ such that $U^{(X,F)}_t=e^{it\widehat{F}}$. From (\ref{ioig5r43}) we see that ${\cal D}_{\mathfrak{K}}$ defined in (\ref{q})
is a $U^{(X,F)}_t$-invariant dense linear subspace of ${\cal H}$. From (\ref{ioig5r43}) it also follows that the functions $\mathbb{R}\ni t \mapsto U_t|\psi\rangle\in {\cal H}$, where $|\psi\rangle\in {\cal D}_{\mathfrak{K}}$, are differentiable. So, see Theorem VII.11  in \cite{R-S}, $ {\cal D}_{\mathfrak{K}}$ is an essential domain of the infinitesimal generator $\widehat{F}$ of $U^{(X,F)}_t$.

Differentiating both sides of the equation (\ref{ioig5r43}) with respect to $t\in\mathbb{R}$ we obtain
\begin{equation}\label{poik}
    i\widehat{F}\mathfrak{K}(p)=\xi (\mathfrak{K})(p)=\left((\pi^*\times \nu^*)^{-1}(X, F)\mathfrak{K}\right)(p)=(H^*(X)\mathfrak{K})(p)+\mathfrak{K}(p) F(p).
\end{equation}
Note here that for any $p\in P$ one has $Ran\;\mathfrak{K}(p)\subset{\cal D}_{\mathfrak{K}}\subset {\cal D}_{\widehat{F}}$. We also note that the last equality in (\ref{poik}) follows from (\ref{iouj87})-(\ref{znik879}) and (\ref{ioig5r43}).

The symmetricity of $\widehat{F}$ on the domain ${\cal D}_{\mathfrak{K}}$ is equivalent to the equation
\begin{equation}\label{yyyty1}
    0=[(H^*(X)\mathfrak{K})(p)+\mathfrak{K}(p) F(p)]^*\mathfrak{K}(q)+\mathfrak{K}^*(p)[(H^*(X)\mathfrak{K})(q)+\mathfrak{K}(q) F(q)],\nonumber
\end{equation}
where $H^*(X)$ is the horizontal lift of $X\in \Gamma^\infty(TM)$ with respect to $\vartheta$.

From (\ref{poik}) and (\ref{ioiqqon45}) we find that the mean values map $\langle \cdot\rangle:\widehat{F}\mapsto \langle \widehat{F}\rangle$ defined by
\begin{equation}\label{i8u7y8ui}
    \langle \widehat{F}\rangle:=(\mathfrak{K}(p)^*\mathfrak{K}(p))^{-1}\mathfrak K^*(p)\widehat{F} \mathfrak K(p)=-iF(p)\nonumber
\end{equation}
is inverse to the quantization $Q:F \mapsto \widehat{F}$ of the classical generator $F$ of the Hamiltonian flow $\tau_t^F$.

Using the isomorphism $I_K:{\cal H}\rightarrow{\cal H}_{K}$ we can represent the  quantum flow $U^{(X,F)}_t$ and its generator $\widehat{F}$ in terms of the Hilbert space ${\cal H}_{K}$. Namely, we have
\begin{equation}\label{jjiji5}
    I_K\circ U^{(X,F)}_t\circ I_K^{-1}=\bar \Sigma_t\nonumber
\end{equation}
and
\begin{equation}\label{jjiji776}
    I_K\circ \widehat{F}\circ I_K^{-1}=\bar Q_{(X,F)},\nonumber
\end{equation}
where $\bar \Sigma_t$ and $\bar Q_{(X,F)}$ are defined in (\ref{09je43je}) and (\ref{ikksi4e}), respectively.

Having in mind the Hilbert spaces isomorphism $I_K:{\cal H}\rightarrow{\cal H}_{K}$, defined in (\ref{qqqwq}), we see that ${\cal D}_{K}$ defined in (\ref{Afr432}) is a common essential domain of the self-adjoint operators $\bar Q_{(X,F)}$ being the generators of the flows $U^{(X,F)}_t$.

According to Definition \ref{quantization7} the problem of quantization of the classical flow $\mathbb R\ni t\mapsto\tau^{(X,F)}_t\in\mbox{Aut}(P,\vartheta)$ which solves the generalized Hamilton equation (\ref{lie54}) reduces to the finding of a coherent state map $\mathfrak{K}:P\rightarrow \mathcal{B}(V,{\cal H})$ satisfying the conditions $\vartheta=\vartheta_{\mathfrak{K}}$ and (\ref{ioig5r43}). Taking into account that  $K:P\times P\rightarrow {\cal B}(V)$ defines $\mathfrak{K}:P\rightarrow \mathcal{B}(V,{\cal H})$ up to a unitary map $U:{\cal H}\rightarrow{\cal H}$ it is reasonable to formulate these conditions in terms of the positive definite kernel:
\begin{equation}\label{44432f}
    \vartheta(p)=\vartheta_{\mathfrak{K}}(p)=K(p,p)^{-1}d_qK(p,q)|_{q=p},
\end{equation}
\begin{equation}\label{der434}
    K(p,q)=K(\tau^{(X,F)}_t(p),\tau^{(X,F)}_t(q)).
\end{equation}

Now, let us describe the quantization conditions (\ref{44432f}) and (\ref{der434}) in terms of local representations  $\mathfrak{K}_\alpha:=
 \mathfrak{K}\circ s_\alpha:{\cal O}_\alpha\rightarrow \mathcal{B}(V,{\cal H})$ and $K_{\bar\alpha\beta}:=\mathfrak{K}_\alpha^*\circ\mathfrak{K}_\beta:{\cal O}_\alpha\times{\cal O}_\beta\rightarrow{\cal B}(V)$ of $\mathfrak{K}:P\rightarrow \mathcal{B}(V,{\cal H})$ and $K:P\times P\rightarrow {\cal B}(V)$.

\begin{proposition}\label{pt7332}
\begin{enumerate}
\item[(i)] The conditions (\ref{44432f}) and (\ref{der434}) written in terms of a local representation $s_\alpha:{\cal O}_\alpha\rightarrow P$ assume the following form
    \begin{equation}\label{io8888n45}
    \vartheta_\alpha(m)=K_{\bar\alpha\alpha}(m,m)^{-1}d_nK_{\bar\alpha\alpha}(m,n)|_{n=m},
\end{equation}
    \begin{equation}\label{kj9mm2}
K_{\bar\alpha\beta}(m,n)=
g_\alpha(t,m)^\dagger K_{\bar\alpha\beta}(\sigma^{(X,F)}_t (m),\sigma^{(X,F)}_t (n))g_\beta(t,n)
\end{equation}
where $\vartheta_\alpha:=s_\alpha^*\vartheta$, the local cocycle $g_\alpha(t,m)$ and the flow $\sigma^{(X,F)}_t:M\rightarrow M$ are related by (\ref{ioi1coc324}), if we put $\tau_t=\tau^{(X,F)}_t$ and $\sigma_t=\sigma^{(X,F)}_t$.
\item[(ii)] The infinitesimal version of the condition (\ref{kj9mm2}) is the following
\begin{equation}\label{568886ioi}
    {\cal X}(K_{\bar\alpha\beta})(m,n)+ \phi_\alpha(m)K_{\bar\alpha\beta}(m,n)+K_{\bar\alpha\beta}(m,n) \phi_\beta(n)=0,
\end{equation}
where ${\cal X}(K_{\bar\alpha\beta})(m,n):=\frac{d}{dt}K_{\bar\alpha\beta}(\sigma^{(X,F)}_t (m),\sigma^{(X,F)}_t (n))|_{t=0}$, i.e. ${\cal X}(m,n)=(X(m),X(n))$, and $\phi_\alpha(m)$, defined in (\ref{ftery656}), satisfy the equation
\begin{equation}\label{eewes3}
    {\cal L}_X\vartheta_\alpha+d\phi_\alpha+[\vartheta_\alpha,\phi_\alpha]=0
\end{equation}
equivalent to the eq. (\ref{hjam8i}).
\end{enumerate}
\end{proposition}{\it \underline{Proof}}

i) The equality (\ref{kj9mm2}) follows from (\ref{ioi1coc324}).

ii) One obtains (\ref{568886ioi}) by differentiating of (\ref{kj9mm2}) with respect to the parameter $t\in\mathbb{R}$ at $t=0$. The condition (\ref{eewes3}) we obtain from (\ref{hjam8i}) using (\ref{ftery656}), (\ref{pre34d}) and (\ref{pre34df}).

\hfill $\Box$

Let us also mention the  transformation formulas
 \begin{equation}\label{qtr5koh}
\mathfrak{K}_\beta(m)=\mathfrak{K}_\alpha(m)g_{\alpha\beta}(m),\nonumber
\end{equation}
and
\begin{equation}\label{qmak8}
    K_{\bar \beta\gamma}(m,n)=g_{\alpha\beta}(m)^{\dagger} K_{\bar \alpha\delta}(m,n)g_{\delta\gamma}(n),\nonumber
\end{equation}
 where  $g_{\alpha\beta}:{\cal O}_\alpha\cap{\cal O}_\beta\rightarrow G$ is the cocycle defined in (\ref{igtu765h}) and $m\in{\cal O}_\alpha\cap{\cal O}_\beta $, $n\in{\cal O}_\gamma\cap{\cal O}_\delta $, between two local representations.

The domain ${\cal D}_{\widehat{F}}$ of the generator $\widehat{F}$ contains ${\cal D}_{\mathfrak{K}}\subset {\cal D}_{\widehat{F}}$ the dense subset ${\cal D}_{\mathfrak{K}}$ defined in (\ref{q}) which consists of $|\psi\rangle=\sum_{j\in J} \mathfrak{K}_{\beta_j}(n_j)v_j\in{\cal D}_{\widehat{F}}$, where $n_j\in{\cal O}_{\beta_j}$, $v_j\in V$ and $J$ is finite. So, the local version of the formula (\ref{poik}) is the following
\begin{equation}\label{nnnny1}
    i\widehat{F}\mathfrak{K}_\beta(n)v=X(\mathfrak{K}_\beta)(n)v+\mathfrak{K}_\beta(n)\phi_\beta(n)v.
\end{equation}
The domains ${\cal D}(Q_{(X,F)})$ and ${\cal D}(\bar Q_{(X,F)})$ of the Kirillov-Kostant-Souriau operators $Q_{(X,F)}:{\cal D}(Q_{(X,F)})\rightarrow\Gamma^\infty(M,\mathbb{V})$ and $\bar Q_{(X,F)}:{\cal D}(\bar Q_{(X,F)})\rightarrow\Gamma^\infty(M,\bar{\mathbb{V}})$ defined in (\ref{ioosi4e}) and (\ref{ikksi4e}), respectively, in the local representation consists of the vectors
$\psi=\sum_{j\in J} K_{\bar \alpha_j\beta}(m_j,\cdot)v_j \in{\cal D}(Q_{(X,F)})$ and
$\bar \psi=\sum_{j\in J} K_{\bar \beta \alpha_j}(\cdot,n_j)v_j\in{\cal D}(\bar Q_{(X,F)})$.

The local version of (\ref{ioosi4e}) and (\ref{ikksi4e}) are
\begin{equation}\label{n716y1}
    Q_{(X,F)}K_{\bar\alpha\beta}(m,\cdot)v=X(K_{\bar\alpha\beta})(m,\cdot)v+
    K_{\bar\alpha\beta}(m,\cdot)\phi_\beta(\cdot)v
\end{equation}
and
\begin{equation}\label{nn66y1}
   \bar Q_{(X,F)}K_{\bar\alpha\beta}(\cdot,n)v=X(K_{\bar\alpha\beta})(\cdot,n)v+
   \phi_\alpha(\cdot)^\dag K_{\bar\alpha\beta}(\cdot,n)v.
\end{equation}

Summing up the above facts, we see that the problem of quantization of classical flow $\tau^{(X,F)}_t$ described by the generalized Hamilton equation (\ref{lie54}) (in local representation by the equations (\ref{hjam8i}) or by (\ref{eewes3})) is reduced to the problem of finding a solution $K_{\bar\alpha\beta}:{\cal O}_\alpha\times{\cal O}_\beta\rightarrow{\cal B}(V)$ of the equations (\ref{io8888n45}) and (\ref{568886ioi}), where $X$, $F_\alpha$ and $\vartheta_\alpha$ are fixed and satisfy (\ref{eewes3}). In general this is a rather hard task. However, it is possible to do this for some particular cases. For this reason see Section \ref{section6}.

\section{Extension and reduction}

It turns out that if $G\subset GL(V,\mathbb{C})$, then having a principal bundle $P(G,\pi,M)$ and a coherent state map $\mathfrak{K}: P \rightarrow B(V, \mathcal{H})$ one can define in a canonical way two other principal bundles $\widetilde{P}(GL(V, \mathbb{C}), \widetilde{\pi}, M)$ and $U(U(V), \pi^u, M)$ over $M$ with the structural groups $GL(V,\mathbb{C})$ and $U(N)$, respectively.  Moreover the coherent state method of quantization of the generalized Hamiltonian flows on $P(G,\pi,M)$ extends uniquely to each of these principal bundles giving the same quantum flows as in the case of $P(G,\pi,M)$.

Indeed, since $G$ is a Lie subgroup of $GL(V,\mathbb{C})$ one can define the $GL(V,\mathbb{C})$-principal bundle $\widetilde{P}(GL(V,\mathbb{C}),\widetilde{\pi},M)$ over $M$ in the following way:
\begin{itemize}
\item[a)] the total space $\widetilde{P}$ is the quotient $\widetilde{P}:=(P\times GL(V,\mathbb{C}))/G$ defined by the action
\begin{equation}\label{ddder41}
    \widetilde{\Phi}_g: P\times GL(V,\mathbb{C})\ni(p,\widetilde{g})\mapsto (pg,g^{-1}\widetilde{g})\in P\times GL(V,\mathbb{C})
\end{equation}
of $G$ on the product $P\times GL(V,\mathbb{C})$;
\item[b)] the bundle projection $\widetilde{\pi}:\widetilde{P}\rightarrow M$ is defined by
\begin{equation}\label{ffrdell}
    \pi([(p,\widetilde{g})]):=\pi(p),\nonumber
\end{equation}
where $[(p,\widetilde{g})]:=\{(pg,g^{-1}\widetilde{g}): g\in G \}\in \widetilde{P}$;
\item[c)] the right action $\widetilde{\kappa}:\widetilde{P}\times GL(V,\mathbb{C})\rightarrow \widetilde{P}$ of $GL(V,\mathbb{C})$ on $\widetilde{P}$ one defines by
\begin{equation}\label{jjjhujh}
\widetilde{\kappa}_{\widetilde{h}}([(p,\widetilde{g})])=[(p,\widetilde{g})]\widetilde{h}:=[(p,\widetilde{g}\widetilde{h})].\nonumber
\end{equation}
\end{itemize}

One has a natural principal bundles morphism $E:P\rightarrow \widetilde{P}$ defined by
\begin{equation}\label{ZP09}
E(p):=[(p,e)],\nonumber
\end{equation}
which covers the identity map $id:M\rightarrow M$ of $M$. Note here that
\begin{equation}\label{gUOr}
    E(pg)=[(pg,e)]=[(p,g)]=[(p,e])g=E(p)g\nonumber
\end{equation}
for $g\in G\subset GL(V,\mathbb{C})$. Hence, according to the Proposition 6.1 in Chapter II of \cite{K-N}, there exists uniquely defined connection $\widetilde{H}^*: T_{\widetilde{\pi}(\widetilde{p})}M\rightarrow T_{\widetilde{p}}\widetilde{P}$ which in the case considered here  is related to the connection $H^*: T_{\pi(p)}M\rightarrow T_{p}P$ on $P$ by
\begin{equation}\label{LRJ87}
    \widetilde{H}^*_{\widetilde{p}}:=T(\widetilde{\kappa}_{\widetilde{g}}\circ E)(p)\circ  H^*_{p},
\end{equation}
where $\widetilde{p}=[(p,\widetilde{g})]\in \widetilde{P}$.

The connection form $\widetilde{\vartheta}\in \Gamma^\infty(\widetilde{P}, T^*\widetilde{P}\otimes {\cal B}(V))$ corresponding to (\ref{LRJ87}) is the following
\begin{equation}\label{AwSa}
    \widetilde{\vartheta}([(p,\widetilde{g})]):= \widetilde{g}^{-1} \vartheta(p)\widetilde{g}+\widetilde{g}^{-1}d\widetilde{g}.\nonumber
\end{equation}

Having the principal bundles morphism $E:P\rightarrow \widetilde{P}$ we can extend a $G$-equivariant coherent state map $\mathfrak{K}:P\rightarrow \mathcal{B}(V,{\cal H})$ as well as an automorphism $\tau\in \mbox{Aut}(P)$ of $P(G,\pi,M)$ to the ones defined on $\widetilde{P}$. These extensions are defined as follows
\begin{equation}\label{aaa4fd}
    \widetilde{\mathfrak{K}}([(p,\widetilde{g})]):=\mathfrak{K}(p)\widetilde{g},\nonumber
\end{equation}
\begin{equation}\label{sssf4s}
    \widetilde{\tau}([(p,\widetilde{g})]):=[(\tau(p),\widetilde{g})].
\end{equation}
The correctness of the above definitions including their independence on the choice of representative $(pg,g^{-1}\widetilde{g})\in[(p,\widetilde{g})]$, where $g\in G$, one can check easily.

If we extend the map $F:P\rightarrow T_eG$, defined in (\ref{sioipr4}), to the map $\widetilde{F}:\widetilde{P}\rightarrow {\cal B}(V)$ by
\begin{equation}\label{lalala3}
    \widetilde{F}([(p,\widetilde{g})]):=\widetilde{g}^{-1}F(p)\widetilde{g},
\end{equation}
then the following relations
\begin{equation}\label{QEWq1}
    (\widetilde{\textbf{D}}\widetilde{F})(\widetilde{p})=\widetilde{g}^{-1}\textbf{D}F(p)\widetilde{g},
\end{equation}
\begin{equation}\label{swe3eQ}
    \widetilde{H}^*_{\widetilde{p}}(X)\llcorner \widetilde{\Omega}_{\widetilde{p}}=
    \widetilde{g}^{-1}(H^*_p(X)\llcorner \Omega_p)\widetilde{g}
\end{equation}
are fulfilled, where $\widetilde{\textbf{D}}$ and $\widetilde{\Omega}$ are the covariant derivative and the curvature form corresponding to $\widetilde{\vartheta}$.

The above allows us to formulate the following proposition.
\begin{proposition}\label{eaSY}
 A generalized Hamiltonian flow $\tau^{(X,F)}_t\in  \mbox{Aut}(P,\vartheta)$ extends by (\ref{sssf4s})
to the flow $\widetilde{\tau}^{(X,F)}_t\in \mbox{Aut}(\widetilde{P},\widetilde{\vartheta})$ and one has the equality $\widetilde{\tau}^{(X,F)}_t=\tau^{(X,\widetilde{F})}_t$ of the flows, where $\widetilde{F}:\widetilde{P}\rightarrow {\cal B}(V)$ is defined in (\ref{lalala3}), i.e. the extension $\widetilde{\tau}^{(X,F)}_t$ is a generalized Hamiltonian flow $\tau^{(X,\widetilde{F})}_t$ on $\widetilde{P}$.
\end{proposition}{\it \underline{Proof}}

It follows from (\ref{QEWq1}) and (\ref{swe3eQ}) that $(X,\widetilde{F})$ satisfies equation
    \begin{equation}\label{gggtr}
    \widetilde{H}^*(X)\llcorner \widetilde{\Omega}+\widetilde{\textbf{D}}\widetilde{F}=0
    \end{equation}
if and only if $(X,F)$ satisfies equation (\ref{lie54}).

\hfill $\Box$

In order to define the principal bundle $U(U(V),\pi^u,M)$ let us note that having a $G$-equivariant coherent state map $\mathfrak{K}:P\rightarrow \mathcal{B}(V,{\cal H})$ we can define the map $\Phi_K: P\times GL(V,\mathbb{C})\rightarrow P\times GL(V,\mathbb{C})$
\begin{equation}\label{dDDDF}
    \Phi_K(p,\widetilde{g}):= (p,K(p,p)^{\frac12}\widetilde{g}),\nonumber
\end{equation}
where $K(p,p)=\mathfrak{K}(p)^*\mathfrak{K}(p)$. This map intertwines
 \begin{equation}\label{fPWed4}
    \Phi^u_g\circ\Phi_K=\Phi_K\circ\widetilde{\Phi}_g
 \end{equation}
 the action (\ref{ddder41}) with the action
 \begin{equation}\label{Gaw3C}
    \Phi^u_g(p,\widetilde{g}):=(pg,c(p,g)^{-1}\widetilde{g}),
\end{equation}
where the $U(V)$-valued cocycle $c:P\times G\rightarrow U(V)$ is defined by
\begin{equation}\label{ddRed}
    c(p,g):=K(p,p)^{\frac12}gK(pg,pg)^{-\frac12}.\nonumber
\end{equation}
The cocycle property $c(p,g)c(pg,h)=c(p,gh)$ and the unitary property $c(p,g)^\dag c(p,g)=1\!\!1_V$ for $c:P\times G\rightarrow U(V)$ follow from (\ref{qaf21b3}).

From (\ref{fPWed4}) we see that $\Phi_K$ defines an automorphism $[\Phi_K ]:\widetilde{P}\rightarrow \widetilde{P}$ of the principal bundle $\widetilde{P}(GL(V,\mathbb{C}),\widetilde{\pi},M)$ which covers the identity map of the base $M$. The inverse $[\Phi_K ]^{-1}=[\Phi_{K^{-1}}]$ of this automorphism transforms $\widetilde{\vartheta}$ and $\widetilde{F}$ in the following way
\begin{equation}
    \vartheta^u([(p,\widetilde{g})]):=([\Phi_{K^{-1}}]^*\widetilde{\vartheta})([(p,\widetilde{g})])=\nonumber
\end{equation}
\begin{equation}\label{33e4r5G}
    =\widetilde{g}^{-1}\left[ K(p,p)^{\frac{1}{2}}\vartheta(p)K(p,p)^{-\frac{1}{2}}+K(p,p)^{\frac{1}{2}}(dK^{-\frac12})(p,p) \right]\widetilde{g}+\widetilde{g}^{-1}d\widetilde{g},
\end{equation}
and
\begin{equation}\label{6LPtkop}
    F^u([(p,\widetilde{g})]):=(\widetilde{F}\circ[\Phi_{K^{-1}}] )([(p,\widetilde{g})]):=
    \widetilde{g}^{-1}K(p,p)^{\frac{1}{2}}F(p)K(p,p)^{-\frac{1}{2}}\widetilde{g},\nonumber
\end{equation}
respectively.

It follows from $c(p,g)\in U(V)$ that the submanifold $P\times U(V)\subset P\times GL(V,\mathbb{C})$ is invariant with respect to the action (\ref{Gaw3C}). Therefore, one can consider the quotient manifold $U:=(P\times U(V))/G$ as the total space of a $U(V)$-principal bundle $U(U(V),\pi^u,M)$ over $M$ with bundle projection map $\pi^u:U\rightarrow M$ and the right action $\kappa^u: U\times U(V)\rightarrow U$ of $U(V)$ defined as follows
\begin{equation}\label{ZMHgy7}
    \pi^u([(p,c)]):=\pi(p),\nonumber
\end{equation}
\begin{equation}\label{gSP3F}
    \kappa^u([(p,\widetilde{c})],c):=[(p,\widetilde{c}c)].\nonumber
\end{equation}
Let us note here that $[(p,\widetilde{c})]\in(P\times U(V))/G$ is defined by $[(p,\widetilde{c})]:=\{(pg,c(p,g)^{-1}\widetilde{c}): g\in G \}$.

Taking the Lie algebra ${\mathfrak U}(V):=\{Y\in{\cal B}(V):
\langle Yv,w\rangle+\langle v,Yw\rangle=0 \;\mbox{for}\;v,w\in V  \}$  of $U(V)$ and the real vector space ${\mathfrak H}(V)$ of the Hermitian $\langle Hv,w\rangle=\langle v,Hw\rangle$ endomorphisms $H\in{\cal B}(V)$ one obtains the Ad$(U(V))$-invariant splitting ${\cal B}(V)={\mathfrak U}(V)\oplus{\mathfrak H}(V)$ of ${\cal B}(V)$. So, according to Proposition 6.4 of Chapter II in \cite{K-N}, the anti-Hermitian part of $\vartheta^u$ after restriction $\vartheta^a:=\frac12(\vartheta^u-(\vartheta^u)^\dag)|_U$ to $U\subset \widetilde{P}$ defines a connection form on the $U(V)$-principal bundle $U(U(V),\pi^u,M)$. Restricting $F^a:=\frac12(F^u-(F^u)^\dag)|_U$ to $U\subset \widetilde{P}$ one obtains a ${\mathfrak U}(V)$-valued $U(V)$-equivariant function on the total space of $U(U(V),\pi^u,M)$.

Therefore, by fixing a coherent state map $\mathfrak{K}:P\rightarrow \mathcal{B}(V,{\cal H})$ on the total space of principal bundle $P(G,\pi,M)$ we reduce a generalized Hamilton system on $\widetilde{P}$ described by (\ref{gggtr}) to the one defined on $U$ by $\vartheta^a$ and $F^a$. Let us stress here that this reduction depends on the choice of $\mathfrak{K}:P\rightarrow \mathcal{B}(V,{\cal H})$ in a canonical way.

Now, we discuss the problem of quantization of the generalized Hamiltonian flows $\tau^{(X,F)}_t$, $\tau^{(X,\widetilde{F})}_t$ and $\tau^{(X,F^a)}_t$ on $P$, $\widetilde{P}$ and $U$, respectively, which are reciprocally related.

We collect the main facts under our consideration in the subsequent proposition.
\begin{proposition}\label{propositiok}
If a flow $\tau^{(X,F)}_t\in \mbox{Aut}(P,\vartheta)$ is quantized in the sense of Definition \ref{quantization7}, then:
\begin{enumerate}
\item[(i)] The  flow $\tau^{(X,\widetilde{F})}_t$ is also quantized in sense of Definition \ref{quantization7}, i.e. one has $\widetilde{\vartheta}=\vartheta_{\widetilde{\mathfrak{K}}}$ and $\widetilde{U}^{(X,F)}_t\widetilde{\mathfrak{K}}([(p,\widetilde{g})])=
    \widetilde{\mathfrak{K}}(\tau^{(X,\widetilde{F})}[(p,\widetilde{g})])$. Additionally one has the equality $\widetilde{U}^{(X,F)}_t=U^{(X,F)}_t$ of the quantum flows.

\item[(ii)] The total space $U$ of $U(U(V),\pi^u,M)$ is invariant with respect to $\tau^{(X,\widetilde{F})}_t\in \mbox{Aut}(\widetilde{P},\widetilde{\vartheta})$ and  $\tau^{(X,\widetilde{F})}_t|_U=\tau^{(X,F^a)}_t\in \mbox{Aut}(U,\vartheta^a)$. The connection form $\vartheta^a$ satisfies
    \begin{equation}\label{iooiqon455}
    \vartheta^a=\vartheta^a_{\mathfrak{a}}:=\mathfrak{a}^*d\mathfrak{a}
     \end{equation}
     the quantization condition (i) of Definition \ref{quantization7}, for the coherent state map $\mathfrak{a}:U\rightarrow\mathcal{U}(V,{\cal H})\subset \mathcal{B}(V,{\cal H})$ defined by
\begin{equation}\label{dq32wderf}
    \mathfrak{a}([(p,\widetilde{c})]):=\mathfrak{K}(p)K(p,p)^{-\frac12}\widetilde{c},
\end{equation}
where $\mathcal{U}(V,{\cal H})$ is the set of partial isometries of Hilbert space $V$ into the Hilbert space ${\cal H}$.
\item[(iii)] The generalized Hamiltonian flow $\tau^{(X,F^a)}_t\in \mbox{Aut}(U,\vartheta^a)$ is quantized in the sense of Definition \ref{quantization7} and the quantum flow $U^{(X,F^a)}_t$ corresponding to it is equal $U^{(X,F^a)}_t=U^{(X,F)}_t$ to the quantum flow  $U^{(X,F)}_t$.
\end{enumerate}
\end{proposition}{\it \underline{Proof}}

(i) From $\vartheta=\vartheta_{\mathfrak{K}}$ we have
\begin{equation}
    \widetilde{\vartheta}([(p,\widetilde{g})])= \widetilde{g}^{-1} (\mathfrak{K}(p)^*\mathfrak{K}(p))^{-1}\mathfrak{K}(p)^*d\mathfrak{K}(p)\widetilde{g}+
    \widetilde{g}^{-1}d\widetilde{g}=\nonumber
\end{equation}
\begin{equation}\label{s4eSAW}
    =(\widetilde{\mathfrak{K}}([(p,\widetilde{g})])^*\widetilde{\mathfrak{K}}([(p,\widetilde{g})]))^{-1}
    \widetilde{\mathfrak{K}}([(p,\widetilde{g})])^*d\widetilde{\mathfrak{K}}([(p,\widetilde{g})])=
    \vartheta_{\widetilde{\mathfrak{K}}}([(p,\widetilde{g})]).\nonumber
\end{equation}
Next, from (\ref{ioig5r43}) and (\ref{sssf4s}) taken for $\tau=\tau^{(X,F)}_t$ we obtain
\begin{equation}\label{lLLpo0}
    U_t\widetilde{\mathfrak{K}}([(p,\widetilde{g})])=U_t\mathfrak{K}(p)\widetilde{g}=
    \mathfrak{K}(\tau^{(X,F)}_t(p))\widetilde{g}=
    \widetilde{\mathfrak{K}}(\tau^{(X,\widetilde{F})}_t([(p,\widetilde{g})])).\nonumber
\end{equation}

(ii) The quantization property for $\tau^{(X,F)}_t$ implies that
\begin{equation}\label{HHDref}
    K(\tau^{(X,F)}_t(p),\tau^{(X,F)}_t(p))=K(p,p).
\end{equation}
This condition is equivalent to the following one
\begin{equation}
    \Phi_K\circ(\tau^{(X,F)}_t\times id)=(\tau^{(X,F)}_t\times id)\circ\Phi_K.\nonumber
\end{equation}
Since the flow $\tau^{(X,F)}_t\times id:P\times GL(V,\mathbb{C})\rightarrow P\times GL(V,\mathbb{C})$ also commutes with the action $\Phi^u:P\times GL(V,\mathbb{C})\times G\rightarrow P\times GL(V,\mathbb{C})$ of $G$ and $\Phi^u_g(P\times U(V))\subset P\times U(V)$ we find that $U\subset \widetilde{P}$ is invariant with respect to the flow $\tau^{(X,\widetilde{F})}_t$. From (\ref{HHDref}) one obtains that $\mathfrak{a}:U\rightarrow\mathcal{U}(V,{\cal H})$ defined in (\ref{dq32wderf}) satisfies
\begin{equation}\label{seW34S}
    U_t\mathfrak{a}([(p,\widetilde{c})])=\mathfrak{a}(\tau^{(X,\widetilde{F})}([(p,\widetilde{c})])).
\end{equation}
Restricting $\vartheta^u=[\Phi_{K^{-1}}]^*\widetilde{\vartheta}=
[\Phi_{K^{-1}}]^*\vartheta_{\widetilde{\mathfrak{K}}}$ to $U\subset \widetilde{P}$ and using (\ref{33e4r5G}) one obtains (\ref{iooiqon455}).

(iii) The quantization of $\tau^{(X,F^a)}_t=\tau^{(X,\widetilde{F})}_t|_U$ as well as the equality $U_t=U^u_t$ follows from (\ref{iooiqon455}) and (\ref{seW34S}).

\hfill $\Box$

According to (\ref{q}) the essential domains of the generators $\widehat{\widetilde{F}}:{\cal D}_{\widetilde{\mathfrak{K}}}\rightarrow{\cal H}$ and $\widehat{F^a}:{\cal D}_{\mathfrak{a}}\rightarrow{\cal H}$ of flows $U_t=e^{it\widehat{\widetilde{F}}}$ and $U_t=e^{it\widehat{F^a}}$ are the following
\begin{equation}
    {\cal D}_{\widetilde{\mathfrak{K}}}:=\left\{ \sum_{i\in {\cal F}} \widetilde{\mathfrak{K}}(\widetilde{p}_i)v_i:\widetilde{p}_i\in \widetilde{P},\;v_i\in V\right\}\nonumber
\end{equation}
and
\begin{equation}
    {\cal D}_\mathfrak{a}:=\left\{ \sum_{i\in {\cal F}} \mathfrak{a}(\widetilde{p}_i)v_i:\widetilde{p}_i\in U,\;v_i\in V\right\},\nonumber
\end{equation}
where $\mathcal{F}$ is a finite subset of $\mathbb{Z}$.

Since $\widetilde{\mathfrak{K}}(\widetilde{p}_i)v_i=\mathfrak{K}(p_i)\widetilde{g}_iv_i$, $\mathfrak{a}(\widetilde{p}_i)v_i=\mathfrak{K}(p_i)K(p_i,p_i)^{-\frac12}\widetilde{c}_iv_i$ and $v_i\in V$ are chosen in an arbitrary way we obtain that ${\cal D}_{\widetilde{\mathfrak{K}}}={\cal D}_{\mathfrak{a}}={\cal D}_{\mathfrak{K}}$. So, we have also the equalities $\widehat{\widetilde{F}}=\widehat{F^a}=\widehat{F}$ for the generators. However, the formula (\ref{poik}) taken for $\widehat{\widetilde{F}}$ and $\widehat{F^a}$ is different from the one for $\widehat{F}$. Namely, we have
\begin{equation}\label{y7y6y78}
    i\widehat{\widetilde{F}}\widetilde{\mathfrak{K}}(\widetilde{p})=
    (H^*(X)\widetilde{\mathfrak{K}})(\widetilde{p})+\widetilde{\mathfrak{K}}(\widetilde{p}) \widetilde{F}(\widetilde{p}),
\end{equation}
where $\widetilde{p}=[(p,\widetilde{g})]\in \widetilde{P}$, and
\begin{equation}\label{mmm6y78}
    i\widehat{F^a}\mathfrak{a}(\widetilde{p})=
    (H^{a*}(X)\mathfrak{a})(\widetilde{p})+\mathfrak{a}(\widetilde{p}) F^a(\widetilde{p}),
\end{equation}
where $\widetilde{p}=[(p,\widetilde{c})]\in U$.

From (\ref{y7y6y78}) and (\ref{mmm6y78}) we easily see that the mean value functions $\langle\widehat{\widetilde{F}}\rangle$ and $\langle\widehat{F^a}\rangle$ for these generators on the coherent states are equal
\begin{equation}\label{jjik78}
    \langle\widehat{{\widetilde{F}}}\rangle([(p,\widetilde{g})])=\widetilde{g}^{-1}\langle\widehat{F}\rangle(p)g
    =\widetilde{F}([(p,\widetilde{g})]),\nonumber
\end{equation}
\begin{equation}\label{mmSSS78}
    \langle\widehat{{F^a}}\rangle([(p,\widetilde{c})])=(\widehat{F}\circ [\Phi_{K^{-1}}])([(p,\widetilde{g})])\nonumber
\end{equation}
to the generators $\widetilde{F}$ and $F^a$ of the Hamiltonian flows $\tau^{(X,\widetilde{F})}_t$ and $\tau^{(X,F^a)}_t$, respectively.

Taking the positive kernels
\begin{equation}\label{KEE44}
    \widetilde{K}(\widetilde{p},\widetilde{q})=\widetilde{\mathfrak{K}}(\widetilde{p})^*\widetilde{\mathfrak{K}}(\widetilde{q})
    =\widetilde{g}^\dag K(p,q) \widetilde{h},\nonumber
\end{equation}
where $\widetilde{p}=[(p,\widetilde{g})],\widetilde{q}=[(q,\widetilde{h})]\in \widetilde{P}$ and
\begin{equation}\label{QQ23r54}
    A(\widetilde{p},\widetilde{q})=\mathfrak{a}(\widetilde{p})^*\mathfrak{a}(\widetilde{q})
    =\widetilde{c}^\dag K(p,p)^{-\frac12}K(p,q)K(q,q)^{-\frac12} \widetilde{b},\nonumber
\end{equation}
where $\widetilde{p}=[(p,\widetilde{c})],\widetilde{q}=[(q,\widetilde{b})]\in U$, we obtain the domains ${\cal D}_{\widetilde{K}}\subset C^\infty_{\overline{GL(V,\mathbb{C})}}(\widetilde{P},V)\cong\Gamma^\infty(M,\bar{\mathbb V})$ and ${\cal D}_{A}\subset C^\infty_{\overline{U(V)}}(U,V)\cong\Gamma^\infty(M,\bar{\mathbb V})$, see (\ref{Afr432}), of the corresponding Kirillov-Kostant-Souriau operators $\bar Q_{(X,\widetilde{F})}:{\cal D}_{\widetilde{K}}\rightarrow \Gamma^\infty(M,\bar{\mathbb V})$ and $\bar Q_{(X,F^a)}:{\cal D}_{A}\rightarrow \Gamma^\infty(M,\bar{\mathbb V})$.

Summing up the above considerations we conclude that there are three equivalent ways of quantizing of the flow $\sigma^{(X,F)}_t$ by the positive kernel method based on the principal bundles $P(G,\pi,M)$, $\widetilde{P}(GL(V,\mathbb{C}),\widetilde{\pi},M)$ and $U(U(V),\pi^u,M)$ over $M$, respectively. The choice of one of these ways depends on the physical as well as mathematical aspects of the model under investigation.

For example, the quantization based on $U(U(V),\pi^u,M)$ is directly related to interpretation of the positive kernel $A:U\times U\rightarrow {\cal B}(V)$ as the matrix valued transition amplitude kernel. More precisely, let us take such $v,w\in V$ that $\|v\|=\|w\|=1$. Then the vectors $\mathfrak{a}(\widetilde{p})v,\mathfrak{a}(\widetilde{q})w\in{\cal H}$ have norm equal to 1 also, i.e. they describe pure states of the system and the transition amplitude between them is $\langle \mathfrak{a}(\widetilde{p})v| \mathfrak{a}(\widetilde{q})w\rangle=\langle v|A(\widetilde{p},\widetilde{q})w \rangle$. So, one can interpret $A(\widetilde{p},\widetilde{q})$ as the transition amplitude matrix between the states $\widetilde{p}$ and $\widetilde{q}$. For more exhaustive discussion of these physical aspects we address to \cite{O1,O2}.

Ending, let us mention that if the base $M$ of the principal bundle $P(G,\pi,M)$ is a complex analytic manifold then it is resonable to use an approach based on the principal bundle  $\widetilde{P}(GL(V,\mathbb{C}),\widetilde{\pi},M)$. As an example of such a situation see Section \ref{section6}.

\section{Quantization of holomorphic flows on non-compact Riemann surfaces}\label{section6}

In this section we will apply the method of quantization presented in Section \ref{section4} to the case when $P(GL(V,\mathbb C),\pi,M)$ is a holomorphic $GL(V, \mathbb{C})$-principal bundle over a non-compact Riemann surface $M$.

There are two reasons which motivated us to consider this case. The first one is its relative simplicity what allows to solve the system of differential equations (\ref{io8888n45}), (\ref{568886ioi})  on  the kernel $K_{\bar\alpha\beta}:{\cal O}_\alpha\times{\cal O}_\beta\rightarrow{\cal B}(V)$ under the assumption that $(X,F_\alpha)\in {\cal P}^\infty_G (P, \vartheta)$. The above assumption, as was shown in Section \ref{section4}, allows us to quantize the flow $\tau^{(X, F)}_t \in \mbox{Aut}(P, \vartheta)$, using the kernel $K_{\bar\alpha\beta}$ obtained in such a way. The second reason is that this type of kernel (equivalently coherent state map) occurs in various problems of quantum optics, e.g. see \cite{H-O-T},\cite{T-O-H-J-Ch}. We omit here the subcase when $M$ is a compact Riemann surface, since then the Hilbert space $\mathcal{H}$ postulated in Definition 4.1. has finite dimension, which makes the theory less interesting from a mathematical point of view, but not necessarily from a physical one, e.g. see \cite{O-W1},   \cite{O-W2}, \cite{O-W3}, \cite{H-Ch-O-T}.

Using the invariants of the flows $\tau^{(X, F)}_t \in \mbox{Aut}(P, \vartheta)$ and the appropriate gauge transformation, we will reduce the equations (\ref{io8888n45}) and  (\ref{568886ioi}) to the linear ordinary differential equation (\ref{MNOgoje}), which is solvable for $(X, F) \in {\cal P}^\infty_G(P, \vartheta)$. The solutions of (\ref{MNOgoje}) are presented through the formula (\ref{strtras4}) in Proposition \ref{ZGF4r} and Proposition \ref{ZGFAAAW}. We will also obtain the integral decompositions (\ref{LPoklpK9}) of the positive kernels $ K_{\bar\beta\beta} (\bar v, z)$ invariant with respect to the flows $\tau^{(X, F)}_t$ on the positive kernels $K_{\bar\beta\beta} (\bar v, z;\lambda)$ presented in Proposition \ref{klopo00ikmj}. The relationship between the $\mathcal{B}(V)$-valued measures $d(\Gamma_0^*E\Gamma_0)(\lambda) $ used for these decompositions and the spectral measure of the generator $\hat{F}$ of the quantum flow $U_t^{(X, F)}= e^{it\hat{F}}$ is described as well.

One proves, eg. see Section III par. 30 in \cite{F}, that for a non-compact Riemann surface $M$ one has $H^1(M,GL(N,{\cal O}))=0$, i.e. any $GL(N,{\cal O})$-valued holomorphic transition 1-cocycle
$(g_{\alpha\beta})\in Z^1(\{{\cal O}\}_{\alpha\in I},GL(N,{\cal O}))$ is solvable $g_{\alpha\beta}=\delta_\alpha\delta_\beta^{-1}$, where the holomorphic maps $\delta_\alpha:{\cal O}_\alpha\rightarrow GL(N,\mathbb C)$, $\alpha\in I$, represent a holomorphic 0-cocycle. Therefore, all holomorphic vector bundles $\mathbb{V}\rightarrow M$ as well as the holomorphic principal bundles $P(GL(V,\mathbb C),\pi,M)$ over $M$ are trivial.

Being in the framework of the above category we will quantize the holomorphic flows $\tau^{(X,F)}_t\in \mbox{Aut}(P,\vartheta)$ only. We will assume also that the coherent state map $\mathfrak{K}:M\times GL(V,\mathbb C)\rightarrow \mathcal{B}(V,{\cal H})$ is a holomorphic map.

The existence of a holomorphic flow $\sigma_t^X=\pi(\tau^{(X,F)}_t)\in \mbox{Aut}(M)$ on a Riemann surface M radically restricts the class of non-compact Riemann surfaces with this property. Namely, see eg. \cite{F-K}, one proves that any non-compact Riemann surface $M$ which admits a non-discrete group of automorphisms is biholomorphic to the one listed below:
\begin{itemize}
\item[(i)] the Gauss plane $\mathbb{C}=\overline{\mathbb{C}}\setminus\{\infty\}$,
\item[(ii)] the punctured Gauss plane $\mathbb{C}^*:= \overline{\mathbb{C}}\setminus\{\infty,0\}$,
\item[(iii)] the unit disc $\mathbb{D}:= \{z\in\mathbb{C}:|z|<1\}$,
\item[(iv)] the punctured unit disc $\mathbb{D}^*:=
\mathbb{D}\setminus \{0\}$,
\item[(v)] an annulus
$\mathbb{A}_r:=\{z\in\mathbb{C}:r<|z|<1\}$, where $0<r<1$.
\end{itemize}
Since the groups $\mbox{Aut}(M)$ of automorphisms of $M=\mathbb{C}, \mathbb{C}^*,\mathbb{D},
\mathbb{D}^*, \mathbb{A}_r\subset \overline{\mathbb{C}}\cong\mathbb{C}\mathbb{P}(1)$ can be considered as the subgroups of $\mbox{Aut}(\mathbb{C}\mathbb{P}(1))\cong SL(2,\mathbb{C})/\mathbb{Z}_2$ we find that:
\begin{itemize}
\item[(i)]
    $\mbox{Aut}(\mathbb{C})=\{z\mapsto \alpha z+\beta: \alpha\in\mathbb{C}^*,
\beta\in\mathbb{C}\}\cong \mathbb{C}^*\ltimes\mathbb{C}$,
\item[(ii)]
    $\mbox{Aut}(\mathbb{C}^*)=\left\{z\mapsto
\alpha z\;\mbox{or}\;z\mapsto \frac{\alpha}{z}: \alpha\in\mathbb{C}^*\right\}\cong
\mathbb{Z}_2\ltimes\mathbb{C}^*$,
\item[(iii)]
    $\mbox{Aut}(\mathbb{D})=\left\{z\mapsto
\frac{\alpha z+\beta}{\overline{\beta}z+\overline{\alpha}}: |\alpha|^2-|\beta|^2=1, \alpha,\beta\in\mathbb{C}\right\}\cong
SU(1,1)/\mathbb{Z}_2$,
\item[(iv)]
    $\mbox{Aut}(\mathbb{D}^*)=\left\{z\mapsto \alpha z:
\alpha\in S^1 \right\}\cong S^1$,
\item[(v)]
    $\mbox{Aut}(\mathbb{A}_r)=\left\{z\mapsto \alpha z\;
\mbox{or}\;z\mapsto r\frac{\alpha}{z}: \alpha\in
S^1\right\}\cong \mathbb{Z}_2\ltimes S^1$.
\end{itemize}

For all these cases $M$ is a circularly symmetric open subset in $\mathbb{C}$. So, the inclusion map $M\ni m\mapsto\iota(m)=:z\in\mathbb{C}$ defines the global chart which is common for all  Riemann surfaces considered here. Hence, the vector field $X\in\Gamma^\infty(TM)$ tangent to a holomorphic flow  $\sigma_t^X\in\mbox{Aut}(M)$ is given by
\begin{equation}\label{bpole1}
X=w(z)\frac{\partial}{\partial z} +
\overline{w(z)}\frac{\partial}{\partial \bar{z}},
\end{equation}
where $w(z)$ is a second order polynomial
\begin{equation}\label{lllkle1}
w(z)=c z^2+a z +b
\end{equation}
with coefficients satisfying the following conditions
\begin{itemize}
\item[(i)] $c=b=0$ if $M=\mathbb{C}^*,\mathbb{D}^*, \mathbb{A}_r$,
\item[(ii)] $c=0$ if $M=\mathbb{C}$,
\item[(iii)] $c=-\overline{b}$ and $a=2i\omega$, $\omega\in\mathbb{R}$, if $M=\mathbb{D}$.
\end{itemize}
From the above conditions we see that if $b=0$ then one also has $c=0$ in (\ref{lllkle1}).

Taking into account the above facts we present below the list of possible holomorphic flows $\sigma_t^X$ on M.

\begin{proposition}\label{propo61q}
The following holomorphic flows are possible:
\begin{itemize}
\item[(i)] for $M=\mathbb{C}^*,\mathbb{D}^*, \mathbb{A}_r$ one has
\begin{equation}\label{ioio12}
    \sigma_t^X(z)=e^{a t}z,
\end{equation}
where $a\in \mathbb{C}\setminus \{0\}$ for $\mathbb{C}^*$ and $a=i\omega$, $\omega\in \mathbb{R}\setminus \{0\}$,  for $\mathbb{D}^*$ and $\mathbb{A}_r$;
 \item[(ii)] for $M=\mathbb{C}$ one has
 \begin{equation}\label{iioio12ty}
    \sigma_t^X(z)=e^{a t}z+\frac{b}{a}(e^{a t}-1),
\end{equation}
where $a\in \mathbb{C}\setminus \{0\}$. Note here that for $a=0$ the formula (\ref{iioio12ty}) reduces to
\begin{equation}\label{nadzisko}
\sigma_t^X(z)=z+b t;
\end{equation}

\item[(iii)] for $M=\mathbb{D}$ one has
\begin{equation}\label{iiogt3}
    \sigma_t^X(z)=\frac{z(\varrho\cosh\varrho t +
i\omega\sinh\varrho t)
+b\sinh\varrho
t}{z\overline{b}\sinh\varrho t
+ \varrho\cosh\varrho t - i\omega\sinh\varrho t},
\end{equation}
if $\varrho:=\sqrt{-\omega^2+|b|^2}\neq0$ and
\begin{equation}\label{iiogyt3}
    \sigma_t^X(z)=\frac{z(1 +
i\omega t)
+b t}{z\overline{b}t
+ 1 - i\omega t},
\end{equation}
if $\varrho=0$. For the formula (\ref{iiogt3}) it is reasonable to distinguish the following two subcases $\varrho\in\mathbb{R}\setminus\{0\}$ $(-\omega^2+|b|^2>0)$ and $\varrho\in i\mathbb{R}\setminus\{0\}$ $(-\omega^2+|b|^2<0)$.
\end{itemize}
\end{proposition}

\begin{remark}
One can consider the flows (\ref{ioio12}-\ref{iiogyt3})  as a one-parameter subgroups of the M\"obius group and in  accordance with the standard classification of its elements we obtain:
\begin{itemize}
\item[(i)] The flows (\ref{ioio12}) are elliptic for $a= i\omega \in i \mathbb{R}$ and loxodromic in all other cases.
\item[(ii)] The flows (\ref{iioio12ty}) are parabolic for $b\neq 0$ and for $b=0$ we obtain (i).
\item[(iii)] The flows (\ref{nadzisko}) are parabolic.
\item[(iv)] The flows (\ref{iiogt3})  are hyperbolic for $\rho^2>0$ and elliptic for $\rho^2<0$.
\item[(v)] The flows (\ref{iiogyt3}) are parabolic.
\end{itemize}
\end{remark}

In order to quantize the flows $\tau^{(X,F)}_t$ whose projections $\pi(\tau^{(X,F)}_t)=\sigma_t^X$ on $M$ are listed in (\ref{ioio12})-(\ref{iiogyt3}), we recall that for every non-compact Riemann surface $M$ the holomorphic principal bundle $P(GL(V,\mathbb C),\pi,M)$ is trivial. So, there exist a holomorphic section $s_\alpha:M\rightarrow P$ and the corresponding trivialization $\mathfrak{K}_\alpha:M\rightarrow \mathcal{B}(V,{\cal H})$ of the coherent state map which are defined on the whole of $M$. The equations (\ref{eewes3}), (\ref{io8888n45}) and (\ref{568886ioi}) in this trivialization assume the following forms
\begin{equation}\label{8u8l98}
    X(F_\alpha)(\bar z,z)+[F_\alpha(\bar z,z),\phi_\alpha(z)]=0,\;\;\;\frac{\partial \phi_\alpha}{\partial \bar{z}}(\bar z,z)=0,
\end{equation}
\begin{equation}\label{op0987}
    \vartheta_\alpha(\overline{z},z)=K_{\bar\alpha\alpha}(\bar z,z)^{-1}\frac{\partial K_{\bar\alpha\alpha}}{\partial z}(\bar z,z)dz,
\end{equation}
\begin{equation}\label{poilot98}
    X(K_{\bar\alpha\alpha})(\bar z,z)+K_{\bar\alpha\alpha}(\bar z,z)\phi_\alpha(z)+\phi_\alpha(z)^\dag K_{\bar\alpha\alpha}(\bar z,z)=0,
\end{equation}
respectively. Note here that the positive kernel $K_{\bar\alpha\alpha}(\bar z,z):=\mathfrak{K}_\alpha (z)^*\mathfrak{K}_\alpha(z)$ is anti-holomorphic in the first variable and holomorphic in the second one.

The relation (\ref{ftery656}) between the classical data $\vartheta_\alpha$, $F_\alpha$, $X$ in this case is the following
\begin{equation}\label{REld3}
    \phi_\alpha(z)=-(X\llcorner\vartheta_\alpha)(\bar z,z)+F_\alpha(\bar z,z).
\end{equation}
Substituting $\vartheta_\alpha(\bar z,z)$ given by (\ref{op0987}), into (\ref{REld3}) we obtain the equation
\begin{equation}\label{LLps4}
    K_{\bar\alpha\alpha}(\bar{z},z)F_\alpha(\bar z,z)=K_{\bar\alpha\alpha}(\bar{z},z)\phi_\alpha(z)+
    w(z)\frac{\partial K_{\bar\alpha\alpha}}{\partial z} (\bar{z},z)
\end{equation}
on the kernel $K_{\bar\alpha\alpha}:M\times M\rightarrow{\cal B}(V)$ complementary to the equation (\ref{poilot98}).

\begin{remark}
Summarizing, we see that the flow $\tau^{(X,F)}_t\in \mbox{Aut}(P,\vartheta)$ can be quantizable iff the kernel $K_{\bar\alpha\alpha}$ satisfies (\ref{op0987}), (\ref{poilot98}) and (\ref{LLps4}) for given classical data $\vartheta_\alpha$, $F_\alpha$, $w$.
\end{remark}

For simplification of the equations (\ref{poilot98}) and (\ref{LLps4}) we make use of the gauge transformations listed in (\ref{igtu765h})--(\ref{ioiju80u}), which in this case are given by
\begin{equation}\label{SSEW7u}
    \vartheta_\beta(\bar{z},z)=g^{-1}_{\alpha\beta}(z)\vartheta_\alpha(\bar{z},z) g_{\alpha\beta}(z)+g^{-1}_{\alpha\beta}(z)(dg_{\alpha\beta})(z),\nonumber
\end{equation}
\begin{equation}\label{HYJ8u5}
    F_\beta(\bar{z},z)=g^{-1}_{\alpha\beta}(z)F_\alpha(\bar{z},z) g_{\alpha\beta}(z),\nonumber
\end{equation}
\begin{equation}\label{dddsed}
    \phi_\beta(z)=g^{-1}_{\alpha\beta}(z)\phi_\alpha(z) g_{\alpha\beta}(z)-w(z)g^{-1}_{\alpha\beta}(z)\frac{\partial g_{\alpha\beta}}{\partial z}(z).
\end{equation}
We see from (\ref{dddsed}) that if the equation
\begin{equation}\label{jjk5ko}
w(z)\frac{\partial
g_{\alpha\beta}}{\partial z}(z)=\phi_\alpha(z)g_{\alpha\beta}(z)
\end{equation}
has a holomorphic solution $g_{\alpha\beta}:M\rightarrow GL(V,\mathbb{C})$ for the given $\phi_\alpha$ and $w$, then there exists a holomorphic section $s_{\beta}:M\rightarrow P$ such that the equations (\ref{poilot98}) and (\ref{LLps4}) reduce to the following ones
\begin{equation}\label{POLot98}
    X(K_{\bar\beta\beta})(\bar z,z)=0,
\end{equation}
\begin{equation}\label{P9Is4}
    K_{\bar\beta\beta}(\bar{z},z)F_\beta(\bar z,z)=
    w(z)\frac{\partial K_{\bar\beta\beta}}{\partial z} (\bar{z},z)
\end{equation}
and equation (\ref{8u8l98}) reduces to
\begin{equation}\label{HHU38}
    X(F_\beta)(\bar z,z)=0.
\end{equation}

The existence of a holomorphic solution $g_{\alpha\beta}\in {\cal O}(M, GL(V,\mathbb{C}))$ of the equation (\ref{jjk5ko}) on $M$ depends on   $\frac1w\phi_\alpha$ which is  a holomorphic function at least on the domain $M_0:=M\setminus\{z_1,z_2\}$, where $z_1,z_2\in M$ are the roots of the polynomial $w$ (the cases $z_1=z_2$ or $\{z_1,z_2\}=\emptyset$ are admissible also).

For the cases mentioned in Proposition \ref{propo61q} one has
\begin{itemize}
\item[(i)]  if $M=\mathbb{C}^*,\mathbb{D}^*, \mathbb{A}_r$ then $M_0=M$;
\item[(ii)] if $M=\mathbb{C}$ then $M_0=\mathbb{C}\setminus \{-\frac ba \}$ for $a\neq 0$ and $M_0= \mathbb{C}$ for $a=0$;
\item[(iii)] if $M=\mathbb{D}$ then $M_0=\mathbb{D}\setminus \{z_1 \}$ for $|b|^2-\omega^2<0$ and  $M_0=\mathbb{D}$ for $|b|^2-\omega^2\geq0$.
\end{itemize}

From the above we have:
\begin{itemize}
\item[(a)] for the cases (ii) and (iii) if the function $\frac1w\phi_\alpha$ extends as a holomorphic function to $\mathbb{C}$ and to $\mathbb{D}$, respectively, then (\ref{jjk5ko}) has a holomorphic solution;
\item[(b)] for the remaining cases we note that the pull-back of the equation (\ref{jjk5ko}) on the universal covering $\widehat{M_0}$ of $M_0$ always has a holomorphic solution and if this solution is invariant with respect to the natural action of the group $Deck(\widehat{M_0}/M_0)$ on $\widehat{M_0}$ then it defines a holomorphic solution of (\ref{jjk5ko}) on $M$.
\end{itemize}

Hence we see that by the gauge transformation (\ref{dddsed}) the large class of functions $\phi_\alpha$ could be brought to $\phi_\beta=0$. Further we will investigate this case only.

Let us note that for $ \phi_\beta=0$ the local form (\ref{hjam8i}) of Hamilton equation (\ref{lie54}) is
\begin{equation}\label{EWEWSes3}
    0={\cal L}_X\vartheta_\beta=X\llcorner d\vartheta_\beta+dF_\beta\nonumber
\end{equation}
where the connection form $\vartheta_\beta(\bar z,z)=\widetilde{\vartheta}_\beta(\bar z,z)dz$ is related by
\begin{equation}\label{assed4rfd}
   F_\beta(\bar z,z)=(X\llcorner\vartheta_\beta)(\bar z,z) =w(z)\widetilde{\vartheta}_\beta(\bar z,z)
\end{equation}
to the function $F_\beta:M\rightarrow{\cal B}(V)$. See equalities (\ref{ftery656}) and (\ref{eewes3}) for this reason. So, in order to consider the holomorphic flow $\tau^{(X,F)}_t$ as a Hamiltonian flow generated by $(X,F)$ we define $\widetilde{\vartheta}_\beta(\bar z,z)$ by (\ref{assed4rfd}). Therefore the kernel $K_{\bar\beta\beta}$ quantizes the Hamiltonian flow $\pi(\tau^{(X,F)}_t)=\sigma^X_t$ if and only if it satisfies the equations (\ref{POLot98}) and (\ref{P9Is4}), while $F_\beta$ satisfies equation (\ref{HHU38}).

Due to the circular symmetry of $M\subset \mathbb C$ we expand $\mathfrak{K}_\beta:M\rightarrow \mathcal{B}(V,{\cal H})$ as a Laurent series
\begin{equation}\label{huhuurxx}
\mathfrak{K}_\beta(z)=\sum_{n\in J}\Gamma_n z^n,
\end{equation}
by definition convergent in the norm topology of the Banach space ${\cal B}(V,{\cal H})$. Let us mention that because  $V$ is finite-dimensional the norm convergence of (\ref{huhuurxx}) is equivalent to its strong convergence. The set of indices $J$, which numerate  $0\neq\Gamma_n\in{\cal B}(V,{\cal H})$ in (\ref{huhuurxx}), is an infinite subset $J\subset \mathbb{Z}$ of the ring of integer numbers $\mathbb{Z}$. For $M=\mathbb{C}, \mathbb{D}$, in particular, one has $J=\mathbb{N}\cup\{0\}$. The condition (\ref{ffke5r}) on the coherent state map   $\mathfrak{K}_\beta:M\rightarrow \mathcal{B}(V,{\cal H})$ implies
\begin{equation}\label{Dcikxx}
 \{\Gamma_nv: n\in J \;\;and\;\; v\in V \}^\perp=\{0\}.
\end{equation}

Taking into account the norm convergence of (\ref{huhuurxx}) we can express
\begin{equation}\label{bCaucdde}
\Gamma_n= \frac{1}{2\pi i}\oint_{S^1_\rho}\frac{\mathfrak{K}_\beta(z)}{z^{n+1}}dz
\end{equation}
the coefficients $\Gamma_n$ by $\mathfrak{K}_\beta:M\rightarrow \mathcal{B}(V,{\cal H})$, where $S^1_\rho:=\{z\in\mathbb{C}:\;|z|=\rho\}\subset M$.

From (\ref{huhuurxx}) we obtain that
\begin{equation}\label{jjijras4}
    K_{\bar\beta\beta}(\bar{z},z)=
    \mathfrak{K}_\beta(z)^*\mathfrak{K}_\beta(z)=\sum_{m,n\in J}\Gamma_m^*\Gamma_n\bar z^m z^n.
\end{equation}
Using this expansion we find that the equation (\ref{POLot98}) is equivalent to the two-variable second order difference equation
\begin{eqnarray}\label{o9orxioid}
 &\overline{c}(m-1)\Gamma_{m-1}^*\Gamma_{n}+c(n-1)\Gamma_{m}^*\Gamma_{n-1}+
 (a n+\overline{a}
 m)\Gamma_m^*\Gamma_{n}  +
  &\nonumber \\ &
 +b(n+1)\Gamma_{m}^*
 \Gamma_{n+1}+\overline{b}
 (m+1)\Gamma_{m+1}^*\Gamma_{n}=0
\end{eqnarray}
on the coefficients $\Gamma_m^*\Gamma_{n}\in {\cal B}(V)$, where one assumes that $\Gamma_{-1}=0$ if $J=\mathbb{N}\cup\{0\}$.

In the next proposition we present some properties of the coefficients $\Gamma_n\in{\cal B}(V,{\cal H})$ whenever they satisfy the equation (\ref{o9orxioid}).

\begin{proposition}\label{ortogxx}
\begin{itemize}
\item[(i)] If $b=c=0$ and $Re\;a\neq0$, then $\Gamma_n=0$ for $n\in J\setminus \{0\}$.
\item[(ii)] If $b=c=0$ and $a=i\omega\in i\mathbb{R}\setminus\{0\}$, then
\begin{equation}\label{bkombxx}
    \Gamma_m^*\Gamma_{n}= \Gamma_n^*\Gamma_{n} \delta_{mn},
    \end{equation}
 for $m,n\in J$.
\item[(iii)] If $b\neq 0$ then $ J=\mathbb{N}\cup\{0\}$ and $\Gamma_n$ are linearly independent in ${\cal B}(V,{\cal H})$.
\end{itemize}
\end{proposition}{\it Proof}:

For $b=c=0$ the equations (\ref{o9orxioid}) reduces to
\begin{equation}\label{baaaxx}
(a n+\overline{a} m)\Gamma_m^*\Gamma_{n}=0
\end{equation}
for all $m,n\in J$. Thus, for $Re\;a \neq0$ one obtains that $\Gamma_n=0$ if $n\neq0$. This proves (i). In the case $Re\;a =0$ the equations (\ref{baaaxx}) gives $\Gamma_m^*\Gamma_{n}= 0$
for $m\neq n$, so (ii) holds.

(iii) If $b \neq 0$ let us assume that operators  $\Gamma_n$, $n\in \mathbb{N}\cup\{0\}$, are
linearly dependent. Then there exists $N\in\mathbb{N}$ such that
\begin{equation}\label{zal13xx}
\Gamma_N=\sum_{n=0}^{N-1}s_n\Gamma_n
\end{equation}
Let us put in (\ref{o9orxioid}) $n=N$ and rewrite this equation as follows
\begin{eqnarray}\label{zal133xx}
& \left[c(m-1)\Gamma_{m-1}+b(m+1)\Gamma_{m+1}\right]^*\Gamma_N=&\nonumber\\
& -\Gamma_m^*\left[c(N-1)\Gamma_{N-1}+(a N+\bar a m)\Gamma_N+b(N+1)\Gamma_{N+1}\right], &
\end{eqnarray}
Next substituting $\Gamma_N$ defined by (\ref{zal13xx}) into the left hand side of (\ref{zal133xx}) and using eq.
(\ref{o9orxioid}) again we find that
\begin{equation}
b(N+1)\Gamma_m^*\Gamma_{N+1}=-\Gamma_m^*\left[(a N+\bar a m)\Gamma_N+c(N-1)\Gamma_{N-1}\right]+\nonumber
\end{equation}
\begin{equation}
+\Gamma_m^*\left[ \sum_{n=0}^{N-1}s_n[c(n-1)\Gamma_{n-1}+(a n+\bar a m)\Gamma_n
+b(n+1)\Gamma_{n+1}]\right]\nonumber
\end{equation}
for arbitrary $m\in \mathbb{N}\cup\{0\}$. Thus, due to (\ref{Dcikxx}), we have
\begin{equation}
b(N+1)\Gamma_{N+1}=-( a N+\bar a m)\Gamma_N-c(N-1)\Gamma_{N-1}+\nonumber
\end{equation}
\begin{equation}
+\sum_{n=0}^{N-1}s_n[c(n-1)\Gamma_{n-1}+( a n+\bar a m)\Gamma_n
+b(n+1)\Gamma_{n+1}],\nonumber
\end{equation}
which means that $\Gamma_{N+1}$ is a linear combination of $\{\Gamma_{0},\ldots,\Gamma_{N-1} \}$ too. Repeating the above procedure we conclude that $\dim {\cal H}\leq N$, which
leads to the contradiction with the assumption that $\dim {\cal H}=\infty$. Ending, let us note that for $b\neq0$ one has $\Gamma_m\neq0$ for all $m\in \mathbb{N}\cup\{0\}$. Since, assuming $\Gamma_m=0$ for certain $m$ we obtain from eq. (\ref{o9orxioid}) that $c(m-1)\Gamma_{m-1}+b(m+1)\Gamma_{m_+1}=0$. It leads to the linear dependence of $\{\Gamma_n\}_{n\in \mathbb{N}\cup\{0\}}$.

\hfill $\Box$

\begin{corollary}
The vector space
\begin{equation}\label{DciLLx}
{\cal D}_\Gamma:=\left\{\sum_{n\in {\cal F}}\Gamma_nv_n: \;v_n\in V\right\}
\end{equation}
is dense in ${\cal H}$, where $\mathcal{F}$ is a finite subset of $\mathbb{Z}$.
\end{corollary}

Let us observe that for $M=\mathbb{C}$ the flows (\ref{ioio12}) and (\ref{iioio12ty}) are conjugated by the translation $T_{\frac ba}(z):=z+\frac ba$. From this and from the point (i) of Proposition \ref{ortogxx} we have:
\begin{corollary}
If $Re\;a\neq0$, then $\mathfrak{K}_\beta(z)=\gamma_0=const$. Therefore, the flows $\sigma_t^X=\pi(\tau^{(X,F)}_t)$ corresponding to this subcase are not quantizable by the coherent state map method.
\end{corollary}
Hence taking the above statement into account we will assume subsequently that $a=i\omega$, where $\omega \in \mathbb{R}$.

In order to find a solution $K_{\bar\beta\beta}(\bar{z},z)$ of the differential equations (\ref{POLot98}) and (\ref{P9Is4})  we note that using (\ref{POLot98}) and (\ref{HHU38}) one can write $K_{\bar\beta\beta}$ and $F_{\beta}$ as the power series of a real variable $I\in\Delta\subset \mathbb{R} $:
\begin{equation}\label{strtras4}
    K_{\bar\beta\beta}(\bar{z},z)=\Phi_\beta(I(\overline{z},z))=\sum_{n\in J}C_n I(\overline{z},z)^n,
\end{equation}
\begin{equation}\label{sed43edA}
    F_\beta(\bar{z},z)=\Psi_\beta(I(\overline{z},z))=\sum_{n\in J}Q_n I(\overline{z},z)^n,
\end{equation}
where $C_n=C_n^\dagger, Q_n\in {\cal B}(V)$. By definition we assume the norm convergence of these power series expansions of the functions $\Phi_\beta:\Delta\rightarrow{\cal B}(V)$ and $\Psi_\beta:\Delta\rightarrow{\cal B}(V)$ defined on the range $I(M)=:\Delta$ of an invariant $I:M\rightarrow\mathbb{R}$:
\begin{equation}\label{n87uy}
X(I)(\bar z,z)=0,
\end{equation}
of the flow  $\sigma_t^X$ tangent to the vector field $X$ defined in (\ref{bpole1}).

Substituting $K_{\bar\beta\beta}$ and $F_\beta$ given by (\ref{strtras4}) and (\ref{sed43edA}) into the equation (\ref{P9Is4}) we reduce this equation to the ordinary linear equation on the function $\Phi_\beta$
\begin{equation}\label{MNOgoje}
    \nu(I)\frac{d}{dI}\Phi_\beta(I)=\Phi_\beta(I)\Psi_\beta(I),
\end{equation}
where $\Psi_\beta$ is defined by $F_\beta$ through the equation (\ref{sed43edA}),
 and the $i\mathbb R$-valued function $\nu:\Delta\rightarrow i\mathbb R$ is defined
\begin{equation}\label{d4d5d2}
    \nu(I(\overline{z},z)):=w(z)\frac{\partial I}{\partial z} (\bar{z},z)\nonumber
\end{equation}
by the invariant $I(\overline{z},z)$.
The correctness of this definition follows from the equation (\ref{n87uy}) and from $[w\frac{\partial}{\partial z},X]=[w\frac{\partial}{\partial z},w\frac{\partial}{\partial z}+\bar w\frac{\partial}{\partial \bar z}]=0$.

Since $\Phi_\beta=\Phi_\beta^\dagger$ and $\nu=-\bar \nu$ we have from (\ref{MNOgoje}) that $\Phi_\beta\Psi_\beta+\Psi_\beta^\dagger\Phi_\beta=0$ which in the case $M=\mathbb{C},\mathbb{D}$ is equivalent to $\sum_{l=0}^k(C_lQ_{k-l}+Q_{k-l}^\dagger C_l)=0$.

Summarizing the above facts we conclude:
\begin{corollary}
The flow $\tau^{(X,F)}_t$ is quantized by the positive kernel $K_{\bar\beta\beta}=\Phi_\beta\circ I$ if and only if the function $\Phi_\beta:\Delta\rightarrow{\cal B}(V)$ is a solution of the equation (\ref{MNOgoje}), where $\Psi_\beta$ and $\nu$ are related to $F_\beta$ and $X$ through (\ref{sed43edA}) and (\ref{n87uy}), respectively.
\end{corollary}

In the next proposition we will present the invariants $I$ and $\nu\circ I$ in correspondence with flows $\sigma_t^X$ listed in Proposition \ref{propo61q}.

\begin{proposition}\label{AAancja}
 \begin{itemize}
 \item[(i)]  For $M=\mathbb{C}^*,\mathbb{D}^*, \mathbb{A}_r$ one has $w(z)=i\omega z$. Thus
 \begin{equation}\label{ivnared}
   I(\overline{z},z)=\bar zz,\;\;\nu(I)=i\omega I,
\end{equation}
and $I(\mathbb{C}^*)=\Delta=]0,\infty[$, $I(\mathbb{D}^*)= \Delta=]0,1[$ or $I(\mathbb{A}_r)= \Delta=]r^2,1[$, respectively.
 \item[(ii)]   For $M=\mathbb{C}$ one has $w(z)=i\omega z+b$. Thus
 \begin{equation}\label{qq32w32wre}
 I(\overline{z},z)=\omega\bar zz+ i\overline{b} z-ib\overline{z},\;\;\nu(I)=i(\omega I+|b|^2),
 \end{equation}
and $I(\mathbb{C})= \Delta=\mathbb{R}$ if $\omega=0$, $I(\mathbb{C})=\Delta=[-\frac{|b|^2}{\omega},\infty[$ if $\omega>0$ and $I(\mathbb{C})= \Delta=]-\infty, -\frac{|b|^2}{\omega}]$ if $\omega<0$.
 \item[(iii)]  For $M=\mathbb{D}$ one has $w(z)=-\bar b z^2+2i\omega z+b$. Thus
 \begin{equation}\label{ostINV0}
I(\overline{z},z)=\frac{2\omega z\overline{z}+
i\overline{b}z-ib\overline{z}}{1-z\overline{z}}, \;\;\nu(I)=i(I^2+2\omega I+|b|^2),
 \end{equation}
 and $I(\mathbb{D})=\Delta=\mathbb{R}$ for $b\neq0$, $I(\mathbb{D})=\Delta=]-\infty,0[$ for $b=0$, $\omega<0$ and $\Delta=]0,\infty[$ for $b=0$, $\omega>0$.
 \end{itemize}
\end{proposition}{\it Proof}

By straightforward verification.

\hfill $\Box$

Now, based on the formulas given in Proposition \ref{AAancja} we will find the dependence of $\Gamma_m^*\Gamma_{n}\in {\cal B}(V)$ on the $C_{n}\in {\cal B}(V)$. This task is equivalent to solving the difference equation (\ref{o9orxioid}) with the $\{C_n\}_{n\in J}$ as initial data, see \eqref{ggt76yhbn}. We will investigate the subcases mentioned in (\ref{ivnared}), (\ref{qq32w32wre}) and (\ref{ostINV0}) separately.
\begin{proposition}\label{ZGF4r}
For the case $b=0$, see (\ref{ivnared}), which concerns with an arbitrary $M=\mathbb{C}, \mathbb{D}, \mathbb{C}^*,\mathbb{D}^*, \mathbb{A}_r$ we have
\begin{equation}\label{5c21k9d}
    \Gamma_m^*\Gamma_{n}= C_m \delta_{mn}
\end{equation}
for $m,n\in J\subset \mathbb{Z}$.
\end{proposition}{\it Proof}

See formula (\ref{bkombxx}) of Proposition \ref{ortogxx}.

\hfill $\Box$

Now, let us shortly discuss the subcases presented in Proposition \ref{ZGF4r}. From (\ref{5c21k9d}) it follows that the Hilbert space ${\cal H}$ can be decomposed
\begin{equation}\label{swQq7yhru}
    {\cal H}=\bigoplus_{n\in J}\Gamma_nV\nonumber
\end{equation}
into the orthogonal $\Gamma_nV\bot\Gamma_mV$, for $n\neq m$, eigenspaces of $\widehat{F}$ The eigenvalue of $\widehat{F}$ corresponding to $\Gamma_nV$ is $n\omega\in \omega J$, so the subset $\omega J\subset \mathbb{R}$ is the spectrum of $\widehat{F}$. We note here that $J\subset \mathbb{Z}$ could be chosen as an arbitrary infinite subset of $\mathbb{Z}$. Thus the spectral decomposition of the operator $\widehat{F}$ is given by
\begin{equation}\label{rozspekolx}
\widehat{F}=\sum_{n\in J}n\omega \hat{P}_n\nonumber
\end{equation}
where $\hat{P}_n$ are the orthogonal
projectors on the eigenspaces $\Gamma_nV\subset \mathcal{H}$. The kernels $K_{\bar\beta\beta}(\bar{v},z)$ for all these subcases are given by the same formula
\begin{equation}\label{Gpo098jkow}
    K_{\bar\beta\beta}(\bar{v},z)=\sum_{n\in J}C_n(\bar{v}z)^n,\nonumber
\end{equation}
where $0<C_n\in {\cal B}(V)$. For $\dim_\mathbb{C}V=1$ this type of kernels was investigated in \cite{O3}, where their relationship with the theory of $q$-special functions was also shown.

For the remaining subcases, i.e. if $b\neq0$ we will obtain the dependence of $\Gamma_m^*\Gamma_{n}$ on the $C_{n}$ comparing the coefficients in front of monomials $\bar v^m z^n$ occurring in the equality
\begin{equation}\label{potr5f4}
    \sum_{m,n=0}^\infty\Gamma_m^*\Gamma_n\bar v^m z^n=\sum_{n=0}^\infty C_n I(\overline{v},z)^n,\nonumber
\end{equation}
valid for arbitrary $v,z\in \mathbb{C},\mathbb{D}$, which follows from (\ref{jjijras4}) and (\ref{strtras4}).

\begin{proposition}\label{ZGFAAAW}
\begin{itemize}
\item[(i)] For the case $M=\mathbb{C}$ and $b\neq0$ we have
\begin{equation}\label{YYYtr54g}
    \Gamma_m^*\Gamma_{n}=\sum_{l=n}^{n+m}\beta^l_{mn}C_{l},
\end{equation}
where
\begin{equation}\label{bre11xxffr}
\beta^l_{mn}=(i\bar b)^{n-m}{m \choose l-n}{l\choose m}\omega^{m+n-l}|b|^{2l-2n},
\end{equation}
if $m\leq n$.
\item[(ii)] For the case $M=\mathbb{D}$ and $b=-\bar c\neq0$ we have
\begin{equation}\label{YYYtr54g1}
    \Gamma_m^*\Gamma_{n}=\sum_{l=n-m}^{n+m}\beta^l_{mn}C_{l},
\end{equation}
where
\begin{eqnarray}\label{bre11gtyh}
\lefteqn{
 \beta^l_{mn}=i^{n-m}
 \sum_{j=0}^m{l-1+j \choose l-1}{l \choose 2l+2j-n-m}\times }\nonumber\\ &&
\times{2l+2j-n-m \choose j+l-n} (2\omega)^{n+m-l-2j}b^{j+l-n} {\bar b}^{j+l-m}
\end{eqnarray}
if $m\leq n$.

The respective formulas for $m>n$ one obtains by conjugation of the ones presented in  (\ref{bre11xxffr}) and (\ref{bre11gtyh}) and transposition of the indices.

\item[(iii)] For both cases described above one has
\begin{equation}\label{ggt76yhbn}
\Gamma_0^*\Gamma_n=(i\bar{b})^n  C_{n}.
\end{equation}
\end{itemize}

\end{proposition}{\it Proof}

By straightforward verification.

\hfill $\Box$

Next proposition describes the action of generator $\widehat{F}$ of the quantum flow $U^{(X, F)}_t = e^{it\hat{F}}$ on the coefficients $\Gamma_n\in{\cal B}(V,{\cal H})$ of the Laurent expansion (\ref{huhuurxx}).
\begin{proposition}\label{LLKoi98io}
The vector space ${\cal D}_\Gamma$ defined in (\ref{DciLLx}) is contained ${\cal D}_\Gamma\subset{\cal D}_{\widehat{F}}$ in the domain of $\widehat{F}$ and one has
\begin{equation}\label{zakaztak3ws}
    i\widehat{F}\Gamma_n= c(n-1)\Gamma_{n-1}+an\Gamma_n+b(n+1)\Gamma_{n+1},
\end{equation}
for $n\in J$, where we assume $\Gamma_{-1}=0$ if $J=\mathbb{N}\cup\{0\}$.
\end{proposition}{\it Proof}

After applying $e^{it\widehat{F}}$ to both sides of equality (\ref{bCaucdde}) we obtain
\begin{equation}\label{gwiazdka4}
     e^{it\widehat{F}}\Gamma_n=e^{it\widehat{F}} \frac{1}{2\pi i}\oint_{S^1_\rho}\frac{1}{z^{n+1}}\mathfrak{K}_\beta(z)dz=\frac{1}{2\pi i}\oint_{S^1_\rho}\frac{1}{z^{n+1}}e^{it\widehat{F}}\mathfrak{K}_\beta(z)dz.
\end{equation}

Since one has $\|\frac{1}{z^{n+1}}e^{it\widehat{F}}\mathfrak{K}_\beta(z) \|\leq \frac{1}{\rho^{n+1}}\sup_{z\in S^1_\rho}\|\mathfrak{K}_\beta(z) \|<\infty$, so due to Lebesgue's dominated convergence theorem the derivative $\frac{d}{dt}|_{t=0}$ at $t=0$ of the right-hand side of (\ref{gwiazdka4}) commutes with the integral over $S^1_\rho$. Thus, we have
\begin{equation}
     i\widehat{F}\Gamma_n=\frac{d}{dt}\left( \frac{1}{2\pi i}\oint_{S^1_\rho}\frac{1}{z^{n+1}}e^{it\widehat{F}}\mathfrak{K}_\beta(z)dz\right)|_{t=0}=\nonumber
\end{equation}
\begin{equation}
     = \frac{1}{2\pi i}\oint_{S^1_\rho}\frac{1}{z^{n+1}}\frac{d}{dt}\left(\mathfrak{K}_\beta(\sigma_t^X(z))\right)|_{t=0}dz=
     \frac{1}{2\pi i}\oint_{S^1_\rho}\frac{1}{z^{n+1}}X(\mathfrak{K}_\beta)(z)dz=\nonumber
\end{equation}
\begin{equation}
     =\frac{1}{2\pi i}\oint_{S^1_\rho}\frac{1}{z^{n+1}}\sum_{l\in J}(clz^{l+1} + alz^l+ blz^{l-1})\Gamma_ldz=\nonumber
\end{equation}
\begin{equation}
     =\frac{1}{2\pi i}\oint_{S^1_\rho}\frac{1}{z^{n+1}}\sum_{l\in J}[c(l-1)\Gamma_{l-1} + al\Gamma_l+ b(l+1)\Gamma_{l+1}]z^ldz=\nonumber
\end{equation}
\begin{equation}\label{trzmiel}
     = c(n-1)\Gamma_{n-1}+an\Gamma_n+b(n+1)\Gamma_{n+1}.
\end{equation}

Hence, using also Stone's Theorem, we find that the rank of $\Gamma_n$ belongs to ${\cal D}_{\widehat{F}}$ and thus (\ref{zakaztak3ws}) is valid. To obtain the successive equalities in (\ref{trzmiel}) we have used the norm convergence of the series (\ref{huhuurxx}).

\hfill $\Box$

The expression (\ref{nnnny1}) for the generator $\widehat{F}:{\cal D}_{\widehat{F}}\rightarrow{\cal H}$ and the expressions  (\ref{n716y1}) and (\ref{nn66y1}) for the
Kirillov-Kostant-Souriau operators $Q_{(X,F)}:{\cal D}(Q_{(X,F)})\rightarrow\Gamma^{hol}(M,\mathbb{V})$ and $\bar Q_{(X,F)}:{\cal D}(\bar Q_{(X,F)})\rightarrow\Gamma^{antihol}(M,\bar{\mathbb{V}})$ in the case under consideration, i.e. when $\phi_\beta=0$, assume the following forms
\begin{equation}\label{LP0nny1}
    i\widehat{F}\mathfrak{K}_\beta(z)v=w(z)\left(\frac{\partial}{\partial z}\mathfrak{K}_\beta\right)(z)v,
\end{equation}
and
\begin{equation}\label{FFt16y1}
    Q_{(X,F)}K_{\bar\beta\beta}(\bar z,\cdot)v=w(z)\left(\frac{\partial}{\partial z}K_{\bar\beta\beta}\right)(\bar z,\cdot)v,\nonumber
\end{equation}
\begin{equation}\label{REd56y1}
   \bar Q_{(X,F)}K_{\bar\beta\beta}(\cdot,z)v=\overline{w(z)}\left(\frac{\partial}{\partial \bar z}K_{\bar\beta\beta}\right)(\cdot,z)v\nonumber
\end{equation}
The essential domains of the above operators are given by
\begin{equation}\label{LP09oMj}
{\cal D}_{\mathfrak{K}}=\left\{\psi=\sum_{j\in {\cal F}} \mathfrak{K}_\beta(z_j)v_j:\;z_j\in M,\;v_j\in V,\right\},\nonumber
\end{equation}
and by
\begin{equation}\label{Lpoi09oMj}
{\cal D}(Q_{(X,F)})=\left\{\psi=\sum_{j\in {\cal F}} K_{\bar\beta\beta}(\bar z_j,\cdot)v_j:\;z_j\in M,\;v_j\in V,\right\},\nonumber
\end{equation}
\begin{equation}\label{6t56yu9oMj}
{\cal D}(\bar Q_{(X,F)})=\left\{\bar \psi=\sum_{j\in {\cal F}} K_{\bar\beta\beta}(\cdot,z_j)v_j:\;z_j\in M,\;v_j\in V,\right\},\nonumber
\end{equation}
respectively, where ${\cal F}$ is a finite subset of $\mathbb{Z}$.

\begin{proposition}\label{prop67jjjhu}
If $b\neq0$, then $\Gamma_0=\mathfrak{K}_\beta(0)\in {\cal B}(V,{\cal H})$ is a generating element for $i\widehat{F}$ in the Hilbert ${\cal B}(V)$-module ${\cal B}(V,{\cal H})$, i.e. the elements $(i\widehat{F})^n\Gamma_0$, where $n\in\mathbb{N}\cup\{0\}$, are linearly independent and they span a linearly dense subspace of ${\cal B}(V,{\cal H})$.
Moreover, one has
\begin{equation}\label{Aswse43e}
    \Gamma_n=K_n(i\widehat{F})\Gamma_0,
\end{equation}
where the polynomials
\begin{equation}\label{AsLPo43e}
    K_n(i\lambda)=\sum_{l=0}^na_l^n(i\lambda)^l\nonumber
\end{equation}
are defined by the recurrence
\begin{equation}\label{LPO0r2oki98j7}
   K_{n+1}(i\lambda)=  \frac{1}{(n+1)b}\left[i\lambda K_n(i\lambda) -na K_n(i\lambda)-(n-1)c K_{n-1}(i\lambda)\right]
\end{equation}
with the initial conditions $K_{-1}(i\lambda)\equiv0$ and $K_{0}(i\lambda)\equiv1$.
\end{proposition}{\it Proof}

The linear dependence between $\Gamma_0, i\widehat{F}\Gamma_0,\ldots,(i\widehat{F})^n\Gamma_0\in{\cal B}(V,{\cal H})$ and $\Gamma_0, \Gamma_1,\ldots,\Gamma_n\in{\cal B}(V,{\cal H})$, where $n\in \mathbb{N}\cup\{0\}$, given by the equations
\begin{equation}
    \Gamma_k=\sum_{l=0}^ka_l^k(i\widehat{F})^k\Gamma_0,  \nonumber
\end{equation}
where $k=0,1,\ldots,n$, is invertible. Hence, and from the linear independence of $\Gamma_n\in{\cal B}(V,{\cal H})$ we conclude that the vectors $(i\widehat{F})^n\Gamma_0\in{\cal B}(V,{\cal H})$, $n\in \mathbb{N}\cup\{0\}$, are linearly independent.

From Proposition \ref{ortogxx} it follows that the operators $\Gamma_n$ span a dense subset of ${\cal B}(V,{\cal H})$, so, $(i\widehat{F})^n\Gamma_0$ span too.

\hfill $\Box$

Now, let us describe the relationship between the coherent state representation (\ref{LP0nny1}) and the spectral representation of the generator $\widehat{F}$ of the quantum flow $U^{(X,F)}_t=e^{it\widehat{F}}$. Therefore, let $E:\mathbb{R}\rightarrow{\cal L}({\cal H})$ denote the resolution of identity or equivalently the spectral measure $E:{\cal B}(\mathbb{R})\rightarrow{\cal L}({\cal H})$ of the self-adjoint operator $\widehat{F}$, i.e. $\psi\in {\cal D}_{\widehat{F}}$ if and only if
\begin{equation}\label{121232eD}
    \int_{\mathbb{R}}\lambda^2 d\langle E\psi|\psi\rangle(\lambda)<\infty\nonumber
\end{equation}
and one has
\begin{equation}\label{rozkspeD}
    \widehat{F}\psi=\int_{\mathbb{R}}\lambda d(E\psi)(\lambda)
\end{equation}
for $\psi\in {\cal D}_{\widehat{F}}$, see Chapter VI \textsection 66 in \cite{A-G} for details. Above by ${\cal L}({\cal H})$ and by ${\cal B}(\mathbb{R})$ we denoted the lattices of orthogonal projections of ${\cal H}$ and Borel subsets of $\mathbb{R}$, respectively.

From Proposition \ref{LLKoi98io} follows that $\Gamma_nV\subset {\cal D}_{\widehat{F}}$, so, using (\ref{Aswse43e}) and (\ref{rozkspeD}) we find that
\begin{equation}\label{A7867876yh3e}
    \Gamma_n=K_n(i\widehat{F})\Gamma_0=\int_{\mathbb{R}}K_n(i\lambda) d(E\Gamma_0)(\lambda).
\end{equation}
Next, substituting $\Gamma_n$ given by (\ref{A7867876yh3e}) into (\ref{ggt76yhbn}) we obtain
\begin{equation}\label{QRfrl098}
    C_n=(i\bar b)^{-n}\Gamma_0^*\Gamma_n=\int_{\mathbb{R}}(i\bar b)^{-n}K_n(i\lambda) d(\Gamma_0^*E\Gamma_0)(\lambda) =\sum_{l=0}^n i^{l-n}(\bar b)^{-n}a_l^n\mu_l,
\end{equation}
where $\mu_l\in{\cal B}(V)$ defined by
\begin{equation}\label{rozkmomkD}
    \mu_n:=\Gamma_0^*\widehat{F}^n\Gamma_0=\int_{\mathbb{R}}\lambda^n d(\Gamma_0^*E\Gamma_0)(\lambda)
\end{equation}
are the moments of the positive ${\cal B}(V)$-valued measure
\begin{equation}\label{roLPlmkD}
     d(\Gamma_0^*E\Gamma_0)(\lambda):=d[(E\Gamma_0)^*(E\Gamma_0)](\lambda).\nonumber
\end{equation}

Summing up we conclude from (\ref{strtras4}) and (\ref{QRfrl098}) that through the Hamburger moment problem defined by (\ref{rozkmomkD}) one obtains the relationship between the positive kernel $K_{\bar\beta\beta}(\bar{v},z)$ and  the resolution of identity $E:\mathbb{R}\rightarrow{\cal L}({\cal H})$ of $\widehat{F}$.

 Let us  define the Hilbert ${\cal B}(V)$-module $L^2(\mathbb{R},d(\Gamma_0^*E\Gamma_0))$ of ${\cal B}(V)$-valued Borel square integrable functions $\gamma:\mathbb{R}\rightarrow{\cal B}(V)$, i.e. such ones that $\langle\gamma;\gamma\rangle_{L^2}\leq M1\!\!1_V$, where $0< M\in \mathbb{R}$, in sense of ${\cal B}(V)$-valued scalar product
\begin{equation}\label{bodvvaluedsca}
    \langle\gamma;\delta\rangle_{L^2}:=\int_{\mathbb{R}}\gamma(\lambda)^* d(\Gamma_0^*E\Gamma_0)(\lambda)\delta(\lambda)\nonumber
\end{equation}
of the square integrable $\mathcal{B}(V)$-valued functions $\gamma,\delta\in L^2(\mathbb{R},d(\Gamma_0^*E\Gamma_0))$. As it follows from (\ref{DciLLx}) and (\ref{Aswse43e}) $\Gamma_0\in{\cal B}(V,{\cal H})$ is a generating element in ${\cal B}(V,{\cal H})$ for $\widehat{F}$, so we have the isomorphism
\begin{equation}\label{i4s5o7m98}
    \mathcal{I}:L^2(\mathbb{R},d(\Gamma_0^*E\Gamma_0))\ni \gamma\stackrel{\sim}{\longrightarrow}
    \int_{\mathbb{R}} d(E\Gamma_0)(\lambda)\gamma(\lambda)=:\Gamma\in {\cal B}(V,{\cal H})\nonumber
\end{equation}
of the defined above Hilbert ${\cal B}(V)$-modules.

Using the isomorphism $\mathcal{I}$ and (\ref{Aswse43e}) we find that
\begin{equation}\label{Afgfgft63e}
    \Gamma_n=\mathcal{I}(K_n(i\cdot)1\!\!1_V)\nonumber
\end{equation}
and, thus
\begin{equation}\label{se43rfgt6543mn}
\mathfrak{K}_\beta(z)=\sum_{n=0}^\infty\Gamma_n z^n=\mathcal{I}( \mathfrak{K}_\beta(z;\cdot)),
\end{equation}
where the function $\mathfrak{K}_\beta(z;\cdot)\in L^2(\mathbb{R},d(\Gamma_0^*E\Gamma_0))$ is defined by the power series
\begin{equation}\label{DDFgtfh76}
    \mathfrak{K}_\beta(z;\lambda):=\left(\sum_{n=0}^\infty K_n(i\lambda)z^n\right)1\!\!1_V,
\end{equation}
convergent in the norm $\|\cdot \|_{L^2}:= \|\langle \cdot , \cdot \rangle_{L^2}\|$, where by $\|\cdot \|$ we denoted the norm on ${\cal B}(V)$. Later on we will see in Proposition \ref{klopo00ikmj} that it is also point-wise convergent.

We note here that the equivariance condition
\begin{equation}\label{Inv43con51}
    \mathfrak{K}_\beta(\sigma^X_t(z))=e^{it\widehat{F}}\mathfrak{K}_\beta(z)\nonumber
\end{equation}
written in terms of $\mathfrak{K}_\beta(z;\lambda)$ assumes the following form
\begin{equation}\label{Inv43cyyyy}
    \mathfrak{K}_\beta(\sigma^X_t(z);\lambda)=e^{it\lambda}\mathfrak{K}_\beta(z;\lambda).\nonumber
\end{equation}

Taking into account (\ref{se43rfgt6543mn}) and that
$\langle \mathfrak{K}_\beta(v);\mathfrak{K}_\beta(z) \rangle=\langle \mathfrak{K}_\beta(v;\cdot));\mathfrak{K}_\beta(z;\cdot))\rangle_{L^2}$
we obtain the integral decomposition
\begin{equation}\label{LPoklpK9}
    K_{\bar\beta\beta}(\bar{v},z)=\int_{\mathbb{R}} K_{\bar\beta\beta}(\bar{v},z;\lambda)d(\Gamma_0^*E\Gamma_0)(\lambda),
\end{equation}
where
\begin{equation}\label{DLKP0tfh76}
    K_{\bar\beta\beta}(\bar{v},z;\lambda)=\mathfrak{K}_\beta(v;\lambda)^\dag\mathfrak{K}_\beta(z;\lambda)
    =\left(\sum_{m,n=0}^\infty \overline{K_m(i\lambda)}K_n(i\lambda)\bar v^m z^n\right)1\!\!1_V,
    \end{equation}
of the kernel $K_{\bar\beta\beta}(\bar{v},z)=\mathfrak{K}_\beta(v)^*\mathfrak{K}_\beta(z)$ invariant with respect to the flow $\sigma^X_t$ quantized by $e^{it\widehat{F}}$.

Combining (\ref{strtras4}) and (\ref{QRfrl098}) we obtain the expression
\begin{equation}\label{Dfff0tfh76}
    K_{\bar\beta\beta}(\bar{v},z;\lambda):=\left(\sum_{n=0}^\infty K_n(i\lambda)\frac{1}{(i\bar b)^n}I(\bar v, z)^n\right)1\!\!1_V
\end{equation}
on the kernel $K_{\bar\beta\beta}(\bar{v},z;\lambda)$ which is  different to (\ref{DLKP0tfh76}).

The equivalence of (\ref{DLKP0tfh76}) and (\ref{Dfff0tfh76}) follows from the equality
\begin{equation}\label{ssse0tfh76}
    \overline{K_m(i\lambda)}K_n(i\lambda)=\sum_{l=L}^{m+n} \frac{1}{(i\bar b)^{l}}\beta^l_{mn}K_l(i\lambda),\nonumber
\end{equation}
valid for the polynomials $K_n(i\lambda)$, where $\beta^l_{mn}$   are given: by (\ref{bre11xxffr}) and $L=n$ for $M=\mathbb{C}$;  by (\ref{bre11gtyh}) and $L=n-m$  for $M=\mathbb{D}$.

Comparing the right-hand sides of (\ref{DDFgtfh76}) and (\ref{Dfff0tfh76}) we obtain
\begin{equation}\label{QWEr45lp0}
    K_{\bar\beta\beta}(\bar{v},z;\lambda)=\mathfrak{K}_\beta\left(\frac{1}{i\bar b}I(\bar v, z);\lambda\right),
\end{equation}
where $I(\bar z, z)$ is the $\sigma^X_t$-invariant presented in (\ref{qq32w32wre}) and (\ref{ostINV0}) of the Proposition \ref{AAancja}. In the next proposition we will present expressions on $K_n(i\lambda)$, $\mathfrak{K}_\beta(z)$ and $K_{\bar\beta\beta}(\bar{v},z;\lambda)$ for the cases when $b\neq0$.

\begin{proposition}\label{klopo00ikmj}
\begin{itemize}
\item[(i)] If $M=\mathbb{C}$ and $b\neq0$ then we have
\begin{equation}\label{sw34e4fr5}
    K_n(i\lambda)=\frac{(-i\omega)^{n-1}}{n!b^n}\left(-\frac{1}{\omega}\lambda\right)_n,
\end{equation}
where $(x)_n=x(x+1)\ldots(x+n-1)$ is the  Pochhammer symbol, 
\begin{equation}\label{Khhh74fr5}
    \mathfrak{K}_\beta(z;\lambda)=  \frac{i}{\omega} \, _1F_0\left(-\frac{1}{\omega}\lambda; -\frac{i\omega}{b}z\right)1\!\!1_V,
\end{equation}
\begin{equation}\label{Khahaha74fr5}
    K_{\bar\beta\beta}(\bar{v},z;\lambda)=  \frac{i}{\omega} \, _1F_0\left(-\frac{1}{\omega}\lambda; -\frac{\omega^2}{|b|^2}\bar vz-\frac{i\omega}{b} z+\frac{i\omega}{\bar b}\overline{v})\right)1\!\!1_V.
\end{equation}
If $a=i\omega=0$ the formulas above take the form
\begin{equation}\label{SSwe4fr5}
    K_n(i\lambda)=\frac{1}{n!}\left(\frac{i\lambda}{b}\right)^n,
\end{equation}
\begin{equation}\label{KLwe4fr5}
    \mathfrak{K}_\beta(z;\lambda)=e^{i\frac{\lambda}{b}z }1\!\!1_V,
\end{equation}
\begin{equation}\label{KPOe4fr5}
    K_{\bar\beta\beta}(\bar{v},z;\lambda)=e^{i\lambda\left(\frac{z}{b} -\frac{\bar v}{\bar b}\right) }1\!\!1_V.
\end{equation}
\item[(ii)] If $M=\mathbb{D}$ and $b\neq0$ then for $|b|^2-\omega^2\neq0$ we find that polynomials $K_n$ are given by (non-orthogonal) Meixner-Pollaczek polynomials $P^{(0)}_n\left(x;\varphi \right)$ as follows
 \begin{equation}
    K_n(i\lambda)=\left(\frac{A }{2ib\sin \varphi}  \right)^n P^{(0)}_n\left(-i\lambda/A;\varphi \right)=\nonumber
\end{equation}
\begin{equation} \label{LKPowe4fr5}
    =\left(\frac{A e^{i\varphi}}{2ib\sin \varphi}  \right)^n\frac{(2\mu)_n}{n!}\;_2F_1\left(-n,\mu-\frac{i\lambda}{A};2\mu;1-e^{-2i\varphi}  \right)|_{\mu=0}
 \end{equation}
 where $A:=-\frac{2\omega}{|\omega|}\sqrt{|b|^2-\omega^2}$ and $\cos\varphi:=\frac{|\omega|}{|b|}$ (if $|b|^2-\omega^2>0$, then $\varphi\in [0,\pi/2]$, if $|b|^2-\omega^2<0$, then $A$ and $\varphi$ are imaginary, $A,\varphi\in i\mathbb{R}$, and $\varphi/i\in[1,\infty[$) and
\begin{equation}\label{GODrwe4fr5}
    \mathfrak{K}_\beta(z;\lambda)=\left(\frac{2ib\sin\varphi-Ae^{-i\varphi}z}{2ib\sin\varphi-Ae^{i\varphi}z}  \right)^{\frac{i\lambda}{A}}1\!\!1_V,
\end{equation}
\begin{equation}\label{DoGwe4fr5}
    K_{\bar\beta\beta}(\bar{v},z;\lambda)=\left(\frac{2|b|^2\sin\varphi+Ae^{-i\varphi}I(\bar v, z)}{2|b|^2\sin\varphi+Ae^{i\varphi}I(\bar v, z)}  \right)^{\frac{i\lambda}{A}}1\!\!1_V,
\end{equation}
where $I(\bar v, z)$ is given in (\ref{ostINV0}).

For $b\neq0$ and $|b|^2-\omega^2=0$ we have that  the polynomials $K_n$ are expressed by (non-orthogonal) Laguerre polynomials $L^{(-1)}_n(x)$ by
\begin{equation}
    K_n(i\lambda)=\left(\frac{\omega}{ib}\right)^n L^{(-1)}_n(\lambda/\omega)=\nonumber
\end{equation}
\begin{equation}\label{DWare4e4fr5}
    =\left(\frac{\omega}{ib}\right)^n\frac{(\alpha+1)_n}{n!}\;_1F_1\left(-n;\alpha+1;\lambda/\omega\right)|_{\alpha=-1},
\end{equation}
and
\begin{equation}\label{HJko9i874fr5}
    \mathfrak{K}_\beta(z;\lambda)=   \exp{\left(\frac{\lambda z}{\omega z -ib}\right)}1\!\!1_V,
\end{equation}
\begin{equation}\label{HJLoLo874fr5}
    K_{\bar\beta\beta}(\bar{v},z;\lambda)= \exp{\left(\frac{\lambda I(\bar v, z)}{\omega I(\bar v, z)+|b|^2}\right)}1\!\!1_V,
\end{equation}
where $I(\bar v, z)$ is given in (\ref{ostINV0}).
\end{itemize}
\end{proposition}{\it Proof}

For $M=\mathbb{C}$ one can put in (\ref{LPO0r2oki98j7}) $c=0$ (see (\ref{lllkle1})). Thus it is easy to check that (\ref{SSwe4fr5}) and (\ref{sw34e4fr5}) are solutions of this recurrence equation for $a=0$ and $a=i\omega\neq0$, respectively. The relations (\ref{KLwe4fr5}), (\ref{KPOe4fr5}), (\ref{Khhh74fr5}) and (\ref{Khahaha74fr5}) follow immediately from (\ref{DDFgtfh76}), (\ref{Dfff0tfh76}) and the definition of  the hypergeometric functions.

For $M=\mathbb{D}$ one has $c=-\bar b$ and $a=2i\omega$. Let us introduce the polynomials $Q_n$ defined by
\begin{equation}\label{pok9TTTy7}
    Q_n(\lambda):=\frac{n! b^n i^n}{A^n}K_n(-iA\lambda),
\end{equation}
where $A\in \mathbb{C}\setminus \{0\}$.
Then (\ref{LPO0r2oki98j7}) takes the form
\begin{equation}\label{qw32oki98j7}
    \lambda Q_n(\lambda)= Q_{n+1}(\lambda)-n\frac{2\omega }{A} Q_n(\lambda)+n(n-1)\frac{|b|^2}{A^2}b Q_{n-1}(\lambda).\nonumber
\end{equation}
This is a three-term recurrence formula on monic polynomials $Q_n$ with the initial conditions $Q_{-1}(\lambda)\equiv0$ and $Q_{0}(\lambda)\equiv1$. By  Favard's Theorem, see e.g. \cite{Ch} Theorem 4.4, the solutions of (\ref{qw32oki98j7}) are non-orthogonal polynomials described in \cite{K-S}. Namely, for $|b|^2-\omega^2\neq0$ and $A=-\frac{2\omega}{|\omega|}\sqrt{|b|^2-\omega^2}$ the polynomials $Q_n$ are the (non-orthogonal) monic Meixner-Pollaczek polynomials $Q_n(\lambda)= \frac{n!}{(2\sin\varphi)^n}P^{(0)}(\lambda;\varphi)$ and for $|b|^2-\omega^2=0$ and $A=-\omega$ the polynomials $Q_n$ are the (non-orthogonal) monic Laguerre polynomials $Q_n(\lambda)= n!(-1)^nL^{(-1)}(\lambda)$. This proves (\ref{LKPowe4fr5}) and (\ref{DWare4e4fr5}).

To prove (\ref{GODrwe4fr5}), (\ref{DoGwe4fr5}), (\ref{HJko9i874fr5}) and (\ref{HJLoLo874fr5}) it is enough to observe that (\ref{DDFgtfh76}) and (\ref{Dfff0tfh76}) is nothing else than the generating function for the family of polynomials $\{K_n \}_{n=0}^\infty $, which for Meixner-Pollaczek and Laguerre polynomials may be found in \cite{K-S} as well.

\hfill $\Box$

The integral decomposition (\ref{LPoklpK9}) taken for the case $\dim_\mathbb{C}V=1, b\neq 0$ and $a=0$ leads to Bochner's Theorem, see e.g. \cite{R-S}, which is one of most important instruments in the operator theory \cite{A-G} as well as the probability theory. So, as a by-product of our method of quantization applied to the case $M=\mathbb{C},\mathbb{D}$ we obtain a family of Bochner type integral decompositions presented in Proposition \ref{klopo00ikmj} for the positive definite kernels invariant with respect to suitable holomorphic flows $\sigma^X_t$ on the Riemann surfaces $\mathbb{C}$ and $\mathbb{D}$ as well as on the ones which are biholomorphic to them.
We stress here that these decompositions are valid for the arbitrary dimension of the Hilbert space $V$.

Now let us discuss in details the case when $\dim_\mathbb{C}V=1$. In this case one has the natural isomorphism ${\cal B}(V,{\cal H})\cong{\cal H}$. Therefore, after applying the Gram-Schmidt orthonormalization procedure to the elements $\widehat{F}^n\Gamma_0\in{\cal H}$, where $n\in \mathbb{N}\cup\{0\}$, which according to Proposition \ref{prop67jjjhu} are linearly independent and span the vector subspace ${\cal D}_\Gamma\subset {\cal D}_{\widehat{F}}\subset {\cal H}$ dense in ${\cal H}$, we obtain the orthonormal basis
\begin{equation}\label{GDW43eJh8}
    |n\rangle:=P_n(\widehat{F})\Gamma_0\in{\cal D}_\Gamma
\end{equation}
in ${\cal H}$.

The polynomials $P_n(\lambda)$ of degree $n$ appearing in (\ref{GDW43eJh8}) are orthogonal with respect to the positive measure $d(\Gamma_0^*E\Gamma_0)(\lambda)$. They satisfy the three term recurrence
\begin{equation}\label{WeWqWeyh834}
    \lambda P_n(\lambda)=b_{n-1}P_{n-1}(\lambda)+a_n P_n(\lambda)+ b_nP_{n+1}(\lambda)\nonumber
\end{equation}
defined by infinite Jacobi matrix J. One can express the coefficients $a_n$ and $b_n$ of this matrix as well as the polynomials $P_n(\lambda)$ in terms of the moments $\mu_n$, see (\ref{rozkmomkD}), of the measure $d(\Gamma_0^*E\Gamma_0)(\lambda)$. For the respective formulas see Chapter I of \cite{A}.

The self-adjoint operator $\widehat{F}$ expressed in the basis (\ref{GDW43eJh8}) assumes the three-diagonal form
\begin{equation}\label{qwRJ87yh834}
    \widehat{F}| n\rangle=b_{n-1}| n-1\rangle+a_n | n\rangle+ b_n| n+1\rangle,
\end{equation}
as well as in the basis $\Gamma_n$, $n\in \mathbb{N}\cup\{0\}$, see (\ref{zakaztak3ws}).

We summarize the facts mentioned above defining
\begin{equation}\label{bopga1}
\Gamma|n\rangle := \Gamma_n,\nonumber
\end{equation}
\begin{equation}\label{bopga1rt}
P(\widehat{F}^n\Gamma_0) := P_n(\widehat{F})\Gamma_0=|n\rangle,\nonumber
\end{equation}
\begin{equation}\label{bopga1frty}
K(\widehat{F}^n\Gamma_0):= K_n(i\widehat{F})\Gamma_0=\Gamma_n\nonumber
\end{equation}
the operators
\begin{equation}\label{Baz}
\begin{array}{ccc}
& {\cal D}_\Gamma &  \\ &P\swarrow \quad\quad\quad\searrow
K &\\ &{\cal D}_\Gamma\stackrel{\Gamma}
{\longrightarrow}{\cal D}_\Gamma
\end{array}\nonumber
\end{equation}
which by definition intertwine the bases $\{\Gamma_n \}_{n=0}^\infty$, $\{\widehat{F}^n\Gamma_0 \}_{n=0}^\infty$, and $\{|n\rangle \}_{n=0}^\infty$ of the Hilbert space ${\cal H}$.

\begin{proposition}\label{cloSe}
The domain ${\cal D}_{\Gamma^*}$ of the operator $\Gamma^*$ adjoint to $\Gamma$ contains ${\cal D}_\Gamma$, which is also the range of  $\Gamma$. Hence ${\cal D}_{\Gamma^*}$ is dense in ${\cal H}$.
\end{proposition}{\it Proof}

For $\varphi\in{\cal H}$ and $\psi=\sum_{n\in {\cal F}}c_n |n\rangle\in {\cal D}_\Gamma$ one has

\begin{equation}\label{bdow1}
|\langle \varphi |\Gamma\psi\rangle|^2 = \left|\sum_{n\in {\cal F}}
c_n \langle\varphi|\Gamma_{n}\rangle\right|^2 \leq \langle
\psi|\psi\rangle \sum_{n\in {\cal F}}|\langle\varphi|\Gamma_{n}\rangle|^2\leq \langle
\psi|\psi\rangle \sum_{n\in J}|\langle\varphi|\Gamma_{n}\rangle|^2  \nonumber
\end{equation}
Thus we see that if $\sum_{n\in J}|\langle\varphi|\Gamma_{n}\rangle|^2<\infty$ then $\varphi\in{\cal D}(\Gamma^*)$, so, we need to prove that
\begin{equation}\label{S55D9}
    \sum_{n\in J}|\langle \Gamma_m |\Gamma_{n}\rangle|^2<\infty
\end{equation}
for any $m\in J$. Let us consider three subcases mentioned in Proposition \ref{ZGF4r} and Proposition \ref{ZGFAAAW} separately.

For the subcase of Proposition \ref{ZGF4r} it follows from (\ref{5c21k9d}) that $\sum_{n\in J}|\langle \Gamma_m |\Gamma_{n}\rangle|^2=|\langle \Gamma_m |\Gamma_{m}\rangle|^2<\infty$.

To prove (\ref{S55D9}) for the subcase (i) of Proposition \ref{ZGFAAAW} where $M=\mathbb{C}$ let us observe that the quantities $\beta^l_{m,n}$, $l=n,\ldots,n+m$, given by (\ref{bre11xxffr}) form, up to the factor $(i{\bar b})^{n}$, a finite family of polynomials of the variable $n$ of degree no greater than $m$ with coefficients depends on $b,\omega, m$. Thus from (\ref{YYYtr54g}) we obtain that for fixed $m\in \mathbb{N}\cup\{0\}$ one has
\begin{equation}\label{BREBRExxffr}
\sqrt[n]{\langle\Gamma_m|\Gamma_{n}\rangle}\leq\sqrt[n]{\sum_{l=n}^{n+m} |\beta^l_{m,n}| |C_{l}|}\xrightarrow[n\rightarrow \infty]{}0\nonumber
\end{equation}
where the last limit follows from the fact that the right hand side of (\ref{strtras4}) is convergent for arbitrary $I\in\mathbb{R}$ (see (\ref{qq32w32wre})), i.e. $\sqrt[n]{|C_n|}\rightarrow0$, and $\sqrt[n]{|\beta^l_{m,n}| }\rightarrow |b|$. Finally, (\ref{S55D9}) holds due to the root test for the convergence of a series.

The proof of (\ref{S55D9}) for the subcase (ii) of Proposition \ref{ZGFAAAW} where $M=\mathbb{D}$ is similar to the previous case. Namely, from (\ref{bre11gtyh}) follows that the quantities $\beta^l_{m,n}$, $l=n-m,\ldots,n+m$ form, up to the factor $(i{\bar b})^{n}$, a finite family of polynomials of the variable $n$ of degree no greater than $3m$ with coefficients depending on $b,\omega, m$. Thus
\begin{equation}\label{HaSe3xxffr}
\sqrt[n]{\langle\Gamma_m|\Gamma_{n}\rangle}\leq|b|\sqrt[n]{\sum_{l=n-m}^{n+m} |\beta^l_{m,n}| |C_{l}|}\xrightarrow[n\rightarrow \infty]{}0\nonumber
\end{equation}
because the right hand side of (\ref{strtras4}) is convergent for arbitrary grates $|I|\in\mathbb{R}$ (see (\ref{ostINV0})), i.e. $\sqrt[n]{|C_n|}\rightarrow0$.

\hfill $\Box$

We see from the above proposition that the assumption of the Theorem VIII.1 in \cite{R-S} are fulfilled and thus we have:
\begin{proposition}
\begin{itemize}
\item[(i)] The adjoint operator $\Gamma^*$ is closed.
\item[(ii)] The operator $\Gamma$ is closable and one has $\bar \Gamma=\Gamma^{**}$, $(\bar \Gamma)^*=\Gamma^*$.
\item[(iii)] The operator ${\bar \Gamma}^*\bar \Gamma=\Gamma^{*}\Gamma^{**}$ defined on the dense domain ${\cal D}_{{\bar \Gamma}^*\bar \Gamma}=\{\psi\in{\cal D}_{\bar \Gamma}: \bar \Gamma\psi\in {\cal D}_{\bar \Gamma^*} \}$ is self-adjoint (see Exercise 45 in Chapter VIII of \cite{R-S}).
\end{itemize}
\end{proposition}

Let us mention an interesting possibility to describe the coherent state map $\mathfrak{K}_\beta:\mathbb{D}\rightarrow{\cal H}$ which quantizes a holomorphic flow $\sigma^X_t:\mathbb{D}\rightarrow\mathbb{D}$ on the disc. Namely, let us define $\mathfrak{K}_0:\mathbb{D}\rightarrow{\cal H}$ by
\begin{equation}\label{juzpo9awe4h}
    \mathfrak{K}_0(z):=\sum_{n=0}^\infty z^n|n\rangle.
\end{equation}
From (\ref{juzpo9awe4h}) and from the closability of $\Gamma:{\cal D}_\Gamma\rightarrow{\cal D}_\Gamma$ we find that
\begin{equation}\label{HHHYU9awe4h}
    \mathfrak{K}_\beta(z)=\bar \Gamma\mathfrak{K}_0(z).\nonumber
\end{equation}
The above allows us to represent
\begin{equation}\label{prAW434ER52}
    K_{\bar\beta\beta}(\bar{v},z)=\sum_{m,n=0}^\infty \langle m|{\bar \Gamma}^*\bar \Gamma |n \rangle\bar v^m z^n\nonumber
\end{equation}
the positive kernel $K_{\bar\beta\beta}$ in terms of the matrix elements $\langle m|{\bar \Gamma}^*\bar \Gamma |n \rangle$ of the positive self-adjoint operator $\Gamma^{*}\Gamma^{**}={\bar \Gamma}^*\bar \Gamma$.

\begin{proposition}\label{prop:611}
 If $\dim V=1$ then for the flow $\sigma_t(z)=e^{a t}z$, $z\in M=\mathbb{C}, \mathbb{D}, \mathbb{C}^*,\mathbb{D}^*, \mathbb{A}_r$ there exists a holomorphic section $s_\beta:M\rightarrow P(GL(1,\mathbb C),\pi,M)$ for which $\phi_\beta(z)=:\phi_0= const$.
\end{proposition}{\it Proof}:

We need to show that when $\dim V=1$ the equation (\ref{dddsed}) has solution $g_{\alpha\beta}:M\rightarrow \mathbb{C}\setminus \{0\}$ for $\phi_\beta(z)= \phi_0$. Let us rewrite this equation in the following form
\begin{equation}\label{DDA4rt6}
    \frac{\partial g_{\alpha\beta}}{\partial z}(z)=-\frac{\phi_\alpha(z)-\phi_0}{a z}g_{\alpha\beta}(z).
\end{equation}

Because each $M$ is a circularly symmetric domain in $\mathbb{C}$, the holomorphic function $\phi_\alpha:M\rightarrow\mathbb{C}$ is globally defined by its Laurent expansion
\begin{equation}\label{fbblaur}
\phi_\alpha(z)=\sum_{n\in J}p_n z^n.\nonumber
\end{equation}
Let us define a holomorphic function on $M$
\begin{equation}\label{ffbblaur}
\psi(z):=\sum_{n\in J\setminus\{0\}}\frac1np_n z^n.\nonumber
\end{equation}
This definition is correct since $\sqrt[n]{n}\rightarrow1$ for $n\rightarrow\infty$.

Because $z\frac{\partial \psi(z)}{\partial z}=\phi(z)-p_0$, then for $\phi_{0}=p_0$ the holomorphic function $g_{\alpha\beta}(z)=e^{\frac{1}{\alpha}\psi(z)}$ is a solution of (\ref{DDA4rt6}) holomorphic on $M$.

\hfill $\Box$

\begin{corollary}
If $Re\, \phi_0\neq0$, then the flows mentioned in Proposition \ref{prop:611} are not quantizable.
\end{corollary}
{\it Proof}:

If $\phi_\beta (z) = \phi_\beta (0) = \phi_0 \neq 0$, then for $b= 0$ the equations (\ref{poilot98}) and (\ref{huhuurxx}) give
$$ [\delta (m+n) + 2 \mu + i \omega (m-n)] \langle \Gamma_m |\Gamma_n \rangle =0,$$
where $\delta := Re\, a$, $\omega:= Im\, a $ and $\mu := Re\, \phi_0$. Hence, we find that $\Gamma_n \neq 0$ iff $n = - \frac{\mu}{\delta}\in \mathbb{Z}$. So, $\mathfrak{K}_\beta (z) = \Gamma_n z^n$ and from defining property of $\mathfrak{K}_\beta :M \rightarrow \mathcal{H}$ it follows that $\dim_\mathbb{C} \mathcal{H} = 1$. The above contradicts the postulates of Definiction \ref{def41}.

\hfill $\Box$

In the next section we will shortly discuss a possible physical applications of the obtained results.

\section{Remarks about physical applications}\label{section7}

In the theory of quantum mechanical systems there are two naturally distinguished ways of representing quantum Hamiltonians. The first one is by Schr\"odinger differential operator having domain in the Hilbert space $L^2(\mathbb{R}^N, d^Nx)$ of square-integrable functions. The second one, called Fock representation, is given by using the creation and annihilation operators which are the weighted shift operators acting in an abstract Hilbert space. The Schr\"odinger approach is used if one defines a quantum system starting from its classical counterpart (Schr\"odinger quantization). The Fock approach is usually applied to systems which do not have the classical equivalents. This for example happens in quantum optics \cite{F-P,G,W-M} and nuclear physics \cite{K-M}, where the annihilation operators describe the quantum amplitudes of distinguished modes of a quantum physical system.

In order to integrate a quantum system, i.e. to obtain its evolution in  time, one needs to find the spectral resolution of the Hamiltonian. This is the main mathematical task of quantum mechanics leading to the spectral representation of a quantum Hamiltonian.

The coherent states representation of the physical system investigated in this paper was initiated by E. Schr\"odinger in 1926 in the paper \cite{Sch} and next was investigated by V. Fock \cite{Fo} and V. Bargmann \cite{B}, and is known in quantum mechanics as the Bargmann-Fock representation. Later it was revitalized in quantum optics by R.J. Glauber \cite{G}. Let us also mention also the contribution of A. Perelomov, see \cite{P}, to this subject, i.e. the construction of coherent state maps through the irreducible representations of Lie groups.

In the papers \cite{O1,O2} a method of quantization of an arbitrary Hamiltonian system based on the notion of coherent state map was proposed and its generalization to the case of an arbitrary $G$-principal bundle we investigated here. Therefore, the illustration of the above method by its application to concrete physical systems is desirable. The coherent state method of quantization of the harmonic oscillator \cite{Sch} is the most known and one can find it also in the textbooks of quantum mechanics. The two cases related to atomic physics crucial from the physical point of view, i.e.  Kepler and MIC-Kepler systems, were quantized by the coherent state map method in \cite{H-O} and \cite{O-S}, respectively. One can find a large class of systems quantizable by the coherent state method in optics \cite{H-Ch-O-T,H-O-T,T-O-H-J-Ch}, where one usually considers a finite number of modes of an electromagnetic field self-interacting through a nonlinear medium \cite{F-P,P-L,W-M}.  In the papers \cite{O-W1,O-W2,O-W3} the classical and quantum reduction procedures were applied to the system of nonlinearly coupled harmonic oscillators (modes) which leads to quantization of the Hamiltonian systems on circularly symmetric surfaces called Kummer shapes \cite{H,O-W1}. This is a case to which one can apply the results obtained in Section \ref{section6}. The detailed discussion of all mentioned cases would require considerable extension of the paper, so we plan to make it the subject of a subsequent publication.

Finally, let us mention our belief that the kernel decomposition (\ref{LPoklpK9}) presented in Proposition \ref{klopo00ikmj}, which generalizes the one considered in Bochner's Theorem to arbitrary noncompact Riemann surfaces, will find applications in probability theory problems.


\end{document}